\documentclass{article}

\usepackage[utf8]{inputenc}
\usepackage{graphicx}
\usepackage{amsfonts}
\usepackage{subcaption}
\usepackage{url}
\usepackage{float}
\usepackage[normalem]{ulem}
\usepackage[labelfont=bf, font=scriptsize]{caption}
\usepackage{placeins}
\usepackage{authblk}
\usepackage{comment}
\usepackage[a4paper, total={6.3in, 10in}]{geometry}
\usepackage[colorlinks=true, citecolor=blue, filecolor=blue, linkcolor=black]{hyperref}
\usepackage{color}
\usepackage{epstopdf}
\usepackage{comment}
\usepackage{multicol}
\usepackage{amsmath} 
\usepackage{color}
\usepackage[utf8]{inputenc} 
\usepackage[T1]{fontenc}    
\usepackage{url}            
\usepackage{booktabs}       
\usepackage{amsfonts}       
\usepackage{nicefrac}       
\usepackage{microtype}      
\usepackage{lipsum}
\usepackage{placeins}
\usepackage{xcolor,colortbl}
\usepackage{amsmath}
\usepackage[numbers,sort&compress]{natbib}
\usepackage{lastpage,fancyhdr,graphicx}
\usepackage{tabularx}
\usepackage{multirow}
\usepackage{hyperref}
\usepackage{lineno}
\hypersetup{     
    urlcolor=black,
    pdftitle={Overleaf Example},
    pdfpagemode=FullScreen,
    }
\title{\textbf{
Influence maximization under limited network information: Seeding high-degree neighbors}}
\author[1,2,*]{Jiamin Ou}
\author[1]{Vincent Buskens}
\author[3]{Arnout Van De Rijt}
\author[2]{Debabrata Panja}
\affil[1]{\small{Department of Sociology, Utrecht University, Padualaan 14, 3584 CC Utrecht, the Netherlands}}
\affil[2]{Department of Information and Computing Sciences, Utrecht University\\ Princetonplein 5, 3584 CC Utrecht, The Netherlands}
\affil[3]{Department of Political and Social Sciences, European University Institute (EUI)\\ San Domenico di Fiesole (FI), Italy}

\affil[*]{j.ou@uu.nl}

\date{}
\begin{document}

\maketitle

\begin{abstract}
The diffusion of information, norms, and practices across a social network can be initiated by compelling a small number of seed individuals to adopt first. 
Strategies proposed in previous work either assume full network information or large degree of control over what information is collected. However, privacy settings on the Internet and high non-response in surveys often severely limit available connectivity information.Here we propose a seeding strategy for scenarios with limited network information: Only the degrees and connections of some random nodes are known. This new strategy is a modification of ``random neighbor sampling'' and seeds the highest-degree neighbors of randomly selected nodes. In simulations of a linear threshold model on a range of synthetic and real-world networks, we find that this new strategy outperforms other seeding strategies, including high-degree seeding and clustered seeding. 
\end{abstract}

\section{Introduction}
\label{sec:1}
Networks provide conduits for the propagation of information \cite{Vosoughi1146}, social norms \cite{doi:10.1056/NEJMsa066082} and pathogens \cite{doi:10.1056/NEJMoa1003176}. For enhancing the efficacy of propagation in a human interaction network, as well as for effectively inhibiting propagation, it is often useful to locate a core set of nodes --- i.e., \emph{seeds} --- whose triggering can maximize the expected total number of nodes that is eventually activated, or, once immunized, would reduce propagation the most \cite{Morone2015, Valente2012}.
 Public health interventions pertaining to vaccination \cite{10.1239/jap/1389370105}, dissemination of health-relevant behaviours \cite{Kim2015, Chin2021}, and viral marketing that target a small set of influencers \cite{doi:10.1509/jmkr.48.3.425} rely on this principle. 
 
Strategies proposed in prior work either assume full network information \cite{Kempe2003, Kempe2015, Morone2015, Zhu2017, Cohen2003, Li2016, Goyal2011, NC_complex} or large degree of control over what network information is collected \cite{7899466, Mihara2015,Wilder2017,Wilder2018, Eckles2019,Stein2017, Erkol2017}. However, privacy settings on the Internet and high non-response in surveys often severely limit available connectivity information. Strategies that require complete or specific network information can then not be practically implemented. This calls for alternative strategies that can also be employed in more barren scenarios commonly confronted by researchers and practitioners, where the majority of nodes and ties is unknown and where control over what information is available is lacking. We thus ask:
How can one effectively seed a contagion when information on the network is limited?

In the present paper we study the effectiveness of various seeding strategies under conditions of limited network information, namely when only the degrees and connections of some random nodes are known. We do so by simulating the standard linear threshold model of diffusion on a range of both empirical and synthetic networks. All seeding strategies we simulate make do with only the limited network information available. For each strategy we assess the ultimate fraction of activated nodes, averaged across simulation runs.

We find that one seeding strategy stands out: Seeding high-degree neighbors. This strategy builds on the friendship paradox that the neighbors of a randomly selected node have higher expected degree than that node itself \cite{10.1239/jap/1389370105, Kim2015, Paluck566, 10.2307/2781907, Chami2017, 20.500.12613/4943}. It samples a subset of nodes, selects one random neighbor of each node, then uses the highest-degree nodes among the set of neighbors as seeds for the contagion process. We find that this strategy by-and-large (with some noteworthy exceptions) achieves greater overall propagation than other strategies, including random neighbor seeding, high-degree seeding and clustered seeding. We demonstrate that the comparative efficacy of the strategy is maintained even when its competitors are advantaged through modest increases in the number of seeds they can draw on. The success of the strategy may be explained by the fact that it achieves a good balance between its seeds being influential while at the same time being scattering broadly. 

The paper is organized as follows. In Sec. \ref{sec1-1} we review the existing theoretical approaches and current practices in seeding with imperfect knowledge of the network. In  Sec. \ref{sec1-2} we formally introduce the high-degree neighbor seeding strategy, which we refer to in short as one-hopHD, as well as other seeding strategies working under similar limits to network information. In Sec. \ref{sec2} we introduce methods and data concerning the underlying networks, contagion models, and the settings for simulations that are used to evaluate the seeding strategies. In Sec. \ref{sec3-1} we discuss the seeding outcomes of one-hopHD and other seeding strategies in three typical synthetic networks and its implications. In Sec. \ref{sec3-2}-\ref{sec3-4} we analyze the simulation results in 
in 10 real-world social networks, with varying sizes from 100 to 7,600, and demonstrate that the one-hopHD strategy outperforms other seeding strategies with similar network information, especially for networks with high degree variances and short path lengths. We end with a conclusion section \ref{sec4}.

\subsection{Related work: existing theoretical approaches and current practices\label{sec1-1}}

Theoretical studies have shown that identifying the core set of nodes under full network information is NP-hard, meaning that a solution is not guaranteed within polynomial time. Hence, approximation algorithms or heuristics have been developed to provide a solution that is not necessarily the best, but still are reasonably effective \cite{Kempe2003, Kempe2015, Morone2015, Zhu2017, Cohen2003, Li2016, Goyal2011, NC_complex}. These generally assume full knowledge of the network, that is, the complete network structure in terms of which node is connected to which others (i.e., the {\it edge list\/}) is known. For example, a recent study demonstrated that a new topological measure for identifying the central individuals best suited for spreading behaviors that requires peer reinforcement outperformed other strategies \cite{NC_complex}. It calculates the set of all possible paths connecting every two nodes, and thus requires that the complete edge list is available \cite{NC_complex}. As on-the-ground seeding experiences are accumulating, progressively more studies recognize that such full network information is often very expensive, if not impossible, to collect \cite{10.1145/2110363.2110394,POMARE2019,doi:10.1068/b3317t,5428686}. Surveys typically exhibit high non-response rates, so that information about social networks is only available for a minority of respondents. The rising concern of data privacy makes complete network information even more difficult to collect, both online and offline. This renders seeding without full network information a question of rising concerns and practical significance.

Thus far, related theoretical studies on seeding strategies with limited network information have appeared under a variety of names, e.g., \textit {seeding with imperfect network information}, \textit {seeding with partial network information}, \textit {seeding in a partially observable network} or \textit {influence maximization problem for unknown networks}. Based on the type of missing network information and the scalability of the envisaged algorithms, these strategies can be put into two broad categories.

The first category concerns approximation algorithms or heuristics developed for full network information to networks with partially disclosed structure \cite{7899466, Mihara2015,Wilder2017,Wilder2018, Eckles2019,Stein2017, Erkol2017}. The  central idea is to efficiently disclose part of the network structure, i.e., the edge list of some nodes (or ``subnetwork'') so that the established algorithms or heuristics for the full network can be used and the seeding outcomes with full network information can be approximated \cite{Stein2017, Erkol2017}. Inspired by degree-based
heuristic algorithm, Mihara \textit{et al.} \cite{7899466, Mihara2015} used a snowball sampling strategy to reveal the network structure and provide seeds. It starts from a small set of nodes, each of which will disclose their degrees and list of connections. Then the node with highest expected degree from the given information is probed again to reveal its degree and list of connections. Such a process is repeated a number of times and the highest degree nodes after all the probing is used as seeds. Similarly, Eckles \textit{et al.} \cite{Eckles2019} chose a set of random nodes and asked them to reveal their connections with a probability. The revealed connections will go through the same process again to reveal more of the network. It results in a partly revealed network that can be used as the input for the algorithms developed for full network information \cite{Goyal2011, Kempe2015}. While letting go of the assumption of full network information, these studies still assume that there is large degree of control over what network information is collected, e.g. through snowball sampling \cite{7899466, Mihara2015} or probabilistic random walks \cite{Eckles2019}, and pose nontrivial demands on available network information \cite{Stein2017,Erkol2017, Eckles2019}. This renders it impossible to use them for seeding in the scenarios where limits on available network information are predetermined.

The second category dives into stochastic seeding with more realistic assumption of network information, which is relevant to our work in this paper. Public health interventions, marketing campaigns and other real-life problems typically have access to very limited network information such as degrees (i.e., how many connections one has) and random connections of some nodes. Under the constraint of such limited information, a few seeding strategies are used. Guided by the intuition that individuals having many connections are likely to be influential, one can choose the highest degree individuals among those whose degrees are known \cite{Kim2015,Chin2021}. Alternatively, one can use a random seeding strategy to blindly seed as many individuals as possible \cite{Akbarpour2017}.

Within the second category, one approach that has received substantial attention in both theoretical studies and real world practices is the one-hop strategy \cite{Kim2015, Chami2017, 10.2307/2781907, Lattanzi2015}. Depending on the context, it has been named ``acquaintance vaccination'' in epidemiology \cite{10.1239/jap/1389370105},  ``nomination'' \cite{Kim2015} or ``social referents'' \cite{Paluck566}. It is inspired by the phenomena discovered by Feld \cite{10.2307/2781907} that individuals are likely to have fewer friends than their friends do, on average. Therefore, by asking a random node to disclose randomly one of its connections, one can reach a node with a higher degree in most cases. Such a phenomena has been verified in general networks with various degree distributions \cite{DBLP}. Compared to random seeding, one-hop does require some network information, but only at a minimum level. In practice, it is easily achieved by asking random individuals to nominate one of their friends. This strategy has received a great deal of attention in the vaccination and immunization studies \cite{DBLP, 7990601, ZHANG2018920}, and is one of the few seeding strategies used to promote public health interventions such as clean water, nutrients \cite{Kim2015}, maternal and child health \cite{20.500.12613/4943} and combat school bullying \cite{Paluck566}. In a field experiment involving 5,773 households in rural Honduras, Kim \textit{et al.} 2015 \cite{Kim2015} found that interventions targeting friends of randomly selected individuals were more effective than random seeding, resulting in higher adoption rates of nutritional interventions and water purification. Using a similar ``friendship nomination'' technique, another experiment to promote behaviour change in maternal and child health is undergoing for 30,000 people, 176 villages in Western Honduras \cite{20.500.12613/4943}. Paluck \textit{et al.} 2016 \cite{Paluck566} used a variant of one-hop strategy to reduce the bullying in schools. Followed by friends nominations, they targeted the students who had been nominated the most times to combat school bullying.

The current study explores a modification of the one-hop strategy \cite{Kim2015, Chami2017, 10.2307/2781907, Lattanzi2015, DBLP, 7990601, ZHANG2018920,20.500.12613/4943}. Specifically, we investigate how the one-hop strategy would perform if we identify higher degree nodes in a better way. In essence, the one-hop strategy adds one extra step to random seeding, which traces back from random nodes to each of their random connections. It has two underlying principles. The primary principle is to target higher degree nodes by random nomination. The second one, which is more hidden, is that it can distribute seeds broadly throughout the network because of its random selection of initial nodes, though to a lesser extent than random seeding. This study investigates how the one-hop strategy performs if its first principle is enhanced. We formalize and evaluate a variant of one-hop seeding strategy (one-hopHD), i.e., high degree seeding of random connections, which is in line with the idea behind the empirical experiment in Paluck \textit{et al.} 2016 \cite{Paluck566}. 

Interestingly, a recent study argues that network information is not really valuable for seeding, because a small increase in the number of random seeds can catch up with the optimal outcomes with full network information anyway \cite{Akbarpour2017}. However, this observation applies to the independent cascade model studied in \cite{Akbarpour2017}, but not to the linear threshold model we study (details in Sec. 2.3). In the independent cascade model, contagions are ``simple'', that is, people need only one exposure from their connections to pick up the concerned attribute (i.e., adopt), such as the spread of knowledge, easily-convincing rumors and pathogens. For many social phenomena, the contagion is arguably ``complex'' \cite{centola2007complex, centola2010spread}: People need multiple exposures from their connections before they adopt. Complex contagion theory argues that the diffusion of many social norms are complex, for which individuals require connection with multiple activated peers (i.e., ``social reinforcement'') before they become activated. Complex contagion applies to most behaviours that are risky, costly or need to fight against habits such as rioting, healthier diet, more regular exercise, and the purchase of new products. In the linear threshold model \cite{Kempe2003, schelling1978micromotives, valente1995network, watts2002simple, young2020individual}, studied in this paper, contagions are complex as they require that a threshold number of network neighbors adopt first before a focal individual adopts. In linear threshold models, in contrast to independent cascade models, limited network information can be useful for improving seeding outcomes also when a small increase in the number of seed is inexpensive \cite{Kim2015, Chami2017, 20.500.12613/4943, Paluck566}.

We explore a number of seeding strategies, finding a particular variant of the one-hop strategy promising: If we add an extra step to the one-hop strategy, that is, we choose the highest degree nodes from random connections (hereafter called ``one-hopHD''), we obtain the best seeding outcome compared to other seeding strategies using similar levels of network information. We find this to be the case across a broad range of empirical and synthetic social networks. The added value of this new strategy remains largely unchanged when the number of seeds by the random strategy or one-hop strategy increases significantly. This new strategy is in line with the idea of an anti-bullying campaign used in US \cite{Paluck566}: students in 56 schools first nominated a few of their friends, and among those who had been nominated, the top 10$\%$ who received the most nominations were chosen as seeds to take a public stance against conflicts in their schools, resulting in a 30$\%$ drop of reported conflict levels. The core of such an anti-bullying strategy is high-degree seeding of random connections. To the best of our knowledge, this has not been formally proposed as a strategy, neither has it been addressed and evaluated despite a promising empirical evidence. 

The effectiveness of the proposed one-hopHD strategy is evaluated through comparison to other seeding strategies with similar assumptions on the availability of information on the network, i.e., only the degrees and connections of some random nodes. We use four strategies for comparison. The first three are the above-mentioned high-degree seeding, conventional one-hop strategy and random seeding. Specifically, high-degree seeding assumes only the degrees of a small percentage of nodes are given and chooses the ones with highest degrees as seeds. The fourth strategy is clustered seeding, based upon limited network information on degrees and connections. Since complex contagion requires multiple exposure to overcome the threshold, clustered seeding initiates the diffusion process by activating a group of directly connected nodes. In this way, other nodes connected to this group of seeds are likely to receive multiple exposure and become activated as well \cite{Centola+2018, 10.1086/521848}. This seeding strategy is realized by activating nodes that are directly connected to the most central node and iterating the same process until the number of seeds have been reached \cite{NC_complex}. Therefore, the fourth strategy will activate the node with highest degree, from nodes with given degrees, and its direct connections. More details of the strategies used for comparison can be found in Sec. \ref{sec1-2}.

\subsection{The proposed seeding strategy and reference strategies \label{sec1-2}}

To formally define the one-hopHD strategy, we first introduce some notation. Let $\mathcal{G}=(\mathcal{V}, \mathcal{E})$ be the underlying network, in which $\mathcal{V}=\{{v}_1,{v}_2,..., {v}_N\}$ is the set of $N$ nodes, and $\mathcal{E}=\{{e}_1,{e}_2,..., {e}_M\}$ is the set of all $M$ edges. The set of seeds is defined as $\mathcal{S}=\{{d}_1,{d}_2,..., {d}_s\}$, which is a subset of $s$ nodes from $\mathcal{V}$ ($\mathcal{S}\subset\mathcal{V}$).  

In the one-hop strategy, one first draws $s$ random nodes from $ \mathcal{V}$ to form a subset $\mathcal{R}=\{\textit{r}_1,\textit{r}_2,..., \textit{r}_s\}$. Each (random) node discloses one of their connections, resulting in a set of random connections $\mathcal{RN}=\{\textit{rn}_1,\textit{rn}_2,..., \textit{rn}_s\}$ which are directly used as seeds ($\mathcal{S}\equiv \mathcal{RN}$). The proposed one-hopHD strategy adds one more step to the conventional one-hop strategy, which is to seed the highest degree nodes from the set of these random connections ($\mathcal{S}\subset\mathcal{RN}$). To make this possible, the number of random connections $\mathcal{RN}$ must be larger than $s$. Considering the possible network information that can be attained by an affordable network survey in both online and offline environments, following the empirical experiment by Paluck \textit{et al.} 2016 \cite{Paluck566}, here we assume maximum $15\%$ of individuals in the network will respond to the survey and disclose their connections. Moreover, instead of revealing a few of their connections as in the experiment \cite{Paluck566}, we assume that the respondents will provide only one connection, i.e., $\mathcal{R}=\mathcal{RN}=15\%$ of $\mathcal{N}$. For all the nodes in $\mathcal{RN}$, we further assume their degrees will be known by another round of survey. Then, from all the nodes in $\mathcal{RN}$, the top $s$ highest degree nodes will be chosen as seeds. In summary, one-hopHD relies on whom the random connections of $15\% \times N$ random nodes are, and the degrees of these random connections. It demands minimum control over what network information is collected, as the first sample on those people who nominated their neighbors can be completely random and only degrees are asked for the nominated friends in the second round. All other properties of the network remain unknown.

To evaluate the effectiveness of one-hopHD, seeding outcomes by other seeding strategies with a similar level of network information should be used for a fair comparison. Note here that the best possible result with full network information is not used as benchmark due to the computational complexity to reach the best solution under complex contagion. Instead, as specified below, we argue that one-hopHD would be an effective strategy if it can be demonstrated to outperform the other possible seeding strategies with similar levels of network information.
\begin{figure*}
    \centering
    \includegraphics[width=1\linewidth]{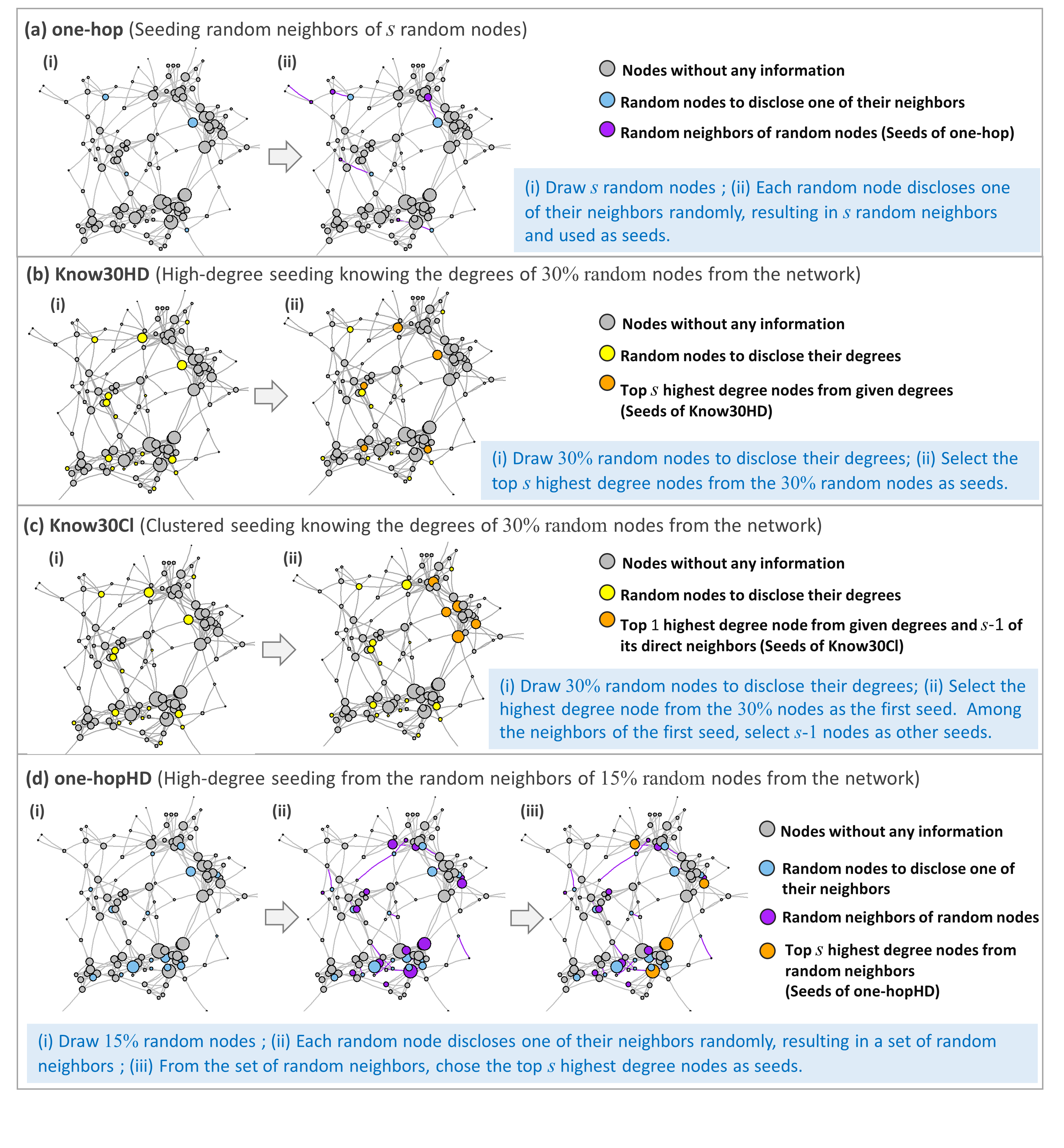}
    \caption{Summary of seed selection strategies and required network information for (a) the one-hop strategy, (b) purely high degree seeding knowing the degrees of $30\%$ random nodes (``Know30HD''), (c) clustered seeding knowing the degrees of $30\%$ random nodes (``Know30Cl''); (d) the one-hopHD strategy which seeds the highest degree nodes from the random connections (``one-hopHD''). Random seeding --- not illustrated here --- will also be used along with one-hop, Know30HD and Know30Cl to evaluate the effectiveness of one-hopHD in synthetic and real social networks under complex contagion.}
    \label{fig1}
\end{figure*}

\begin{itemize}
\item[(i)] The one-hopHD strategy is more efficient than the conventional one-hop strategy with an identical seed set size (Fig. \ref{fig1}a, labeled ``one-hop'').
\item[(ii)]  The one-hopHD strategy is more efficient than the conventional one-hop strategy even when the later can seed 50\% more (``one-hop1.5'').
\item[(iii)] The one-hopHD strategy is more efficient than purely high-degree seeding with an identical seed set size. Specifically, the purely high-degree seeding is based upon partial network information, i.e., knowing the degrees of $30\%$ random nodes. It then chooses the top $s$ highest degree nodes among the $30\%$ nodes with given degree as seeds (Fig. \ref{fig1}b, labeled ``Know30HD'').
\item[(iv)] The one-hopHD strategy is more efficient than clustered seeding with an identical seed set size. Specifically, the clustered seeding is based upon partial network information, i.e., knowing the degrees of $30\%$ random nodes. It then chooses the top highest degree node among the $30\%$ nodes whose degrees are known as the first seed. Among the connections of the first seed, $s-1$ nodes will be chosen randomly to complete the seed set (Fig. \ref{fig1}c, labeled  ``Know30Cl'').
\item[(v)] The one-hopHD strategy is more efficient than random seeding with an identical seed set size (``Random'').
\item[(vi)] The one-hopHD strategy is more efficient than random seeding even if the later can seed twice as many (``Random2'').
\end{itemize}

It is obvious from the above that random and one-hop seeding require less network information than one-hopHD. We have therefore introduced two extra comparisons ``one-hop1.5'' and ``Random2'' in (ii) and (vi) respectively. Moreover, we have introduced Know30HD and Know30Cl to reflect other seeding options with similar levels of network information. While one-hopHD adds an extra step to conventional one-hop by searching the higher degree nodes from random connections, one can argue that seeds can be simply chosen from the nodes with given degrees, without searching for the random connections. We therefore compare one-hopHD to purely high-degree seeding. Given the fact that one-hopHD requires the information in terms of the random connections and the degrees of $15\%$ of the nodes, we assume that Know30HD can know the degrees of $30\%$ of the nodes (``Know30HD''). In other words, Know30HD has more access to the degree information as a compensation to the fact that it does not require any information of direct connections. If one-hopHD outperforms Know30HD, it suggests that the combination of searching random connections and high-degree seeding is more efficient than purely high-degree seeding, even though the later strategy has knowledge on the degrees of more nodes at its disposal. In addition, one-hopHD is compared to Know30Cl since Know30Cl represents another possibility to use the knowledge on the degrees of $30\%$ of the nodes. 

\section{Method and Data\label{sec2}}

\subsection{Networks for simulation and their properties\label{sec2-1}}

First we use three typical synthetic networks: an Erdős–Rényi (ER) a random network, small-world (SW) network and a scale-free (SF) network for studying the performance of seeding strategies under simplified network structures. Albeit not realistic, they capture some properties of real-world social networks. Testing on these synthetic networks can therefore shed light on the performances of seeding strategies when some of the real-world network properties are pushed to the limit (which is hard to test using real-world networks alone). Specifically, the ER random network is characterized by short path lengths, low clustering and moderate degree variances following a normal distribution. It is generated by the Erdős–Rényi–Gilbert model \cite{doi:10.1080/10618600.2012.738106}, in which every two nodes are connected with the same probability (we use 0.1).  The SW network reproduces the short path length and high clustering of real-world networks, but with a relatively homogeneous degree distribution. The SW network is built from a regular lattice, followed by a rewiring probability (we use 0.025) to create shortcuts to mimic the small-world effect \cite{watts_collective_1998}. The SF network differs from the ER and SW networks with a highly heterogeneous degree distribution that follow a power-law decay, extremely low clustering coefficient and long path lengths. It is generated from the Barabási–Albert (BA) model \cite{Albert:2002:rmp, albert2000error} with linear preferential attachment. All the synthetic networks are of the same size (${N}=100$).

We also use ten real-world social networks of various sizes (ranging from 100 to 7,600 nodes) and properties. Among them, 3 are offline social networks with less than 250 nodes, from the SocioPatterns dataset (http://www.sociopatterns.org). They include a friendship network between high school students \cite{10.1371/journal.pone.0136497}, a connection network between the children and teachers in a primary school \cite{10.1371/journal.pone.0023176}, and a daily connection network collected from a Science Gallery \cite{ISELLA2011166}. The other seven networks are online social networks containing 1,000 to 7,600 nodes, such as those of Facebook \cite{NIPS2012_7a614fd0}, Wikipedia \cite{leskovec2010signed, Leskovec2010PredictingPA}, Bitcoin \cite{kumar2016edge, kumar2018rev2} and others. Most of these networks have short path lengths, but differ in other network properties such as clustering coefficient and modularity. Descriptions of the networks and their properties can be found in Table \ref{tab1}.

\begin{table}[H]
\resizebox{\textwidth}{!}{%
    \centering
    \begin{tabular}{l|lcccc}
        \hline
        \textbf{Network name} & \hfill\textbf{Description}\hfill & \textbf{\begin{tabular}{@{}c@{}} Size \\ $(N)$ \end{tabular}} & \textbf{
        \begin{tabular}{@{}l@{}} Avg. shortest \\ path length ($D$)$^a$ \end{tabular}} & \textbf{\begin{tabular}{@{}c@{}} Clustering coefficient \\ ($CC$) \end{tabular}} & \textbf{\begin{tabular}{@{}c@{}} Modularity$^b$ \\ $(M)$ \end{tabular}} \\
        \hline
        high school & \begin{tabular}{@{}l@{}} Connections and friendship relations between students \\ in a high school \cite{10.1371/journal.pone.0136497} \end{tabular}
          & 128 & 4.0 (2.4) & 0.48 &  0.74 \\
        primary school &  \begin{tabular}{@{}l@{}} Connections between the children and \\ teachers in a primary school \cite{10.1371/journal.pone.0023176} \end{tabular} & 236 & 1.8 (1.3) & 0.43 &  0.34 \\
        Infect &  Daily connection networks at a Science Gallery \cite{ISELLA2011166}  & 221 & 3.5 (2.2) & 0.30 &  0.83 \\
        Facebook & Ego-based network of Facebook users \cite{NIPS2012_7a614fd0} & 4039 & 3.7 (2.5) & 0.52 &  0.81 \\
        Bitcoin & Bitcoin Alpha web of trust network \cite{kumar2016edge,kumar2018rev2}  & 3782 & 3.5 (2.2) & 0.08 &  0.42 \\
        API &  Social network of LastFM users from Asia \cite{feather}  & 7623 & 5.2 (3.5) & 0.18 &  0.77 \\
        Wiki & Wikipedia who-votes-on-whom network \cite{Leskovec2010PredictingPA, leskovec2010signed}  & 7066 & 3.5 $(1.9)$ & 0.13 &  0.37 \\
        P2P &  Gnutella peer to peer network \cite{10.1145/1217299.1217301, article} & 6301 & 4.6 (3.2) & 0.02 &  0.34 \\
        CA &  \begin{tabular}{@{}l@{}} Collaboration network from the arXiv covering \\authors from General Relativity and Quantum \\ Cosmology category \cite{10.1145/1217299.1217301} \end{tabular} & 4158 & 6.0 (4.5) & 0.60 &  0.79 \\
        Email &  \begin{tabular}{@{}l@{}} Email network about all incoming and  \\ outgoing email between members from a large \\ European research institution \cite{10.1145/1217299.1217301, 10.1145/3097983.3098069} \end{tabular} & 1006 & 2.6 (1.5) & 0.27 &  0.37 \\
        \hline
    \end{tabular}}
    \caption{Details of the 10 social networks from real world. Notes: $^a$Number in the parenthesis indicates the average path length between the top 10$\%$ highest degree nodes; $^b$Modularity is calculated by the algorithm of Community structure via short random walks (cluster walktrap) using the igraph package in R.} \label{tab1}
\end{table}

\subsection{The linear threshold model of network diffusion\label{sec2-2}}

We study a commonly employed model of network diffusion, the \textit{Linear Threshold Model}. In this model, each node has a state value $v$, which is either $0$ (\textit{inactive}) or $1$ (\textit{active}). All nodes except the seeds start from the inactive state, and once activated, they remain so in the following time steps. At each time step, every node checks the weighted state value of their connections, which is the weighted percentage of activate connections. Assuming that all connections (node $j$) of node $i$ have the same weight for node $i$, the weighted state value of its connections ${w}_i$ at time $t$ is
\begin{eqnarray}
\textit{w}_i(t) &=& \frac{\text {Number of activated connections of node $i$ at time $t$}} {\text {Total number of connections of node $i$}},
\end{eqnarray}
Node \textit{i} will be activated at time $t+1$ only if \(\textit{w}_i\) is equal to or exceeds the predefined threshold  \(\theta_i\), i.e.,
\begin{eqnarray}
    v_i(t+1)= 
\begin{cases}
    1,& \text{if } \textit{w}_i(t)\geq \theta_i\\
    0,              & \text{otherwise.}
\end{cases}
\end{eqnarray}

All the nodes in the network will adopt the same threshold \cite{Morries2000,Berger2001,Chin2021}, but we will vary the value of threshold for the whole network to capture different contagiousness (more details in next section). In our simulations, the total number of activated nodes ${AN}(t)$ at every (discrete) time step $t$, is recorded. The diffusion process is stopped when the number of activated nodes at the (final) time step $t_f$ is equal to the previous time step $t_f-1$. The fraction of activated nodes at time $t_f$ is then the \textit{final activated fraction} $AF(t_f)$, given by
\begin{eqnarray}
AF(t_f) &=& \frac{1} {N} \sum_{i=1}^{N} v_i(t_f),
\end{eqnarray}
and is used as the index for evaluating the effectiveness of seeding strategies.

\subsection{Simulation setup\label{sec2-3}}

We choose the synthetic and real networks introduced in Sec. \ref{sec2-1}, and apply the simulation procedures described in Table \ref{tab2}. Three further conditions need to be specified before starting the simulations. The first one is the number of seeds, $s$. Typical values of $3\% \times N$, $5\% \times N$ and $7\% \times N$ will be used. In most cases, the seed set sizes of different seeding strategies will be the same. Exceptions are Random2 and one-hop1.5, in which the seed set sizes will be larger than the one-hopHD. 

The second condition is of course the seeding strategy. According to the given seeding strategies, seeds will be chosen accordingly in line with the procedures in Table \ref{tab2}. Note that each seeding strategy has its own randomness in seed selection; for instance, for a given network and a given $s$, the seeds provided by the random seeding strategy is different for each run. This is also the case for one-hop, Know30HD, Know30Cl and one-hopHD, which all start from a set of random nodes. To capture the randomness of each strategy, the seed selection process is performed 10,000 times for a given network and $s$. In other words, under each combination of network, seed set size and seeding strategy, a seed matrix with a dimension of  $10,000 \times s$ is generated and used as inputs for the diffusion model. Details in the generation of seed matrix can be found in Table \ref{tab2}. 

The third condition is to define the threshold of the diffusion model. For a given threshold $\theta$, each node in the network is assigned with the same threshold value. We test a wide range of threshold values ranging from 0.15 to 0.6, with an increment of 0.05, for each network and seeding strategy. Once the above three conditions are defined, the diffusion process will be simulated.

\begin{table}[H]
\begin{tabular}{ |p{1cm} p{14cm}|  }
 \hline
 \multicolumn{2}{|l|}{\textbf{Simulation Procedures}} \\
 \hline
 \multicolumn{2}{|l|}{1. Choose an underlying network.} \\
 \hline
 \multicolumn{2}{|l|}{2. Prepare the seed sets for a given size of $s$.} \\
 \hline
  \multicolumn{2}{|l|}{ \textbf{If} \textit{ seeding strategy= ``Random'',}} \\
    & (i) Randomly select $s$ nodes to form a set of random nodes as \(\mathcal{R}\);\\
    & (ii) Record the nodes in \(\mathcal{R}\) as seeds;\\ 
    & (iii) Repeat (i)-(ii) for 10,000 times to generate a seed matrix with a dimension of 10,000\(\times\)$s$.\\ 
  \multicolumn{2}{|l|}{ \textbf{If} \textit{ seeding strategy= ``one-hop'',}} \\
    & (i) Randomly select \emph{s} nodes to form a set of random nodes as \(\mathcal{R}\); \\
    & (ii) Each node in \(\mathcal{R}\) discloses one of their connections to form a set of random connections as  \(\mathcal{RN}\);\\ 
    & (iii) Record the nodes in \(\mathcal{RN}\) as seeds;\\ 
    & (iv) Repeat (i)-(iii) for 10,000 times to generate a seed matrix with a dimension of 10,000\(\times\)$s$.\\ 
  \multicolumn{2}{|l|}{ \textbf{If} \textit{ seeding strategy= ``Know30HD'',}} \\
    & (i) Randomly select 30\%\(\times\)\textit{N} nodes to form a set of random nodes as \(\mathcal{R}\); \\
    & (ii) Each node in \(\mathcal{R}\) discloses its degree;\\ 
    & (iii) Select the top \emph{s} highest degree nodes from \(\mathcal{R}\) as seeds;\\
    & (iv) Repeat (i)-(iii) for 10,000 times to generate a seed matrix with a dimension of 10,000\(\times\)$s$.\\
  \multicolumn{2}{|l|}{ \textbf{If} \textit{ seeding strategy= ``Know30Cl'',}} \\
    & (i) Randomly select 30\%\(\times\)\textit{N} nodes to form a set of random nodes as \(\mathcal{R}\); \\
    & (ii) Each node in \(\mathcal{R}\) discloses its degree;\\ 
    & (iii) Select the highest degree node from \(\mathcal{R}\) as first seed \({d}_1\);\\ 
    & (iv) Seed \(d_1\) randomly discloses $s-1$ of its connections and used as remaining seeds: \(d_2,\ldots,d_s\);\\ 
    & (v) Repeat (i)-(iv) for 10,000 times to generate a seed matrix with a dimension of 10,000\(\times\)$s$.\\ 
   \multicolumn{2}{|l|}{\textbf{If}  \textit{ seeding strategy= ``one-hopHD'',}} \\
    & (i) Randomly select 15\%\(\times\)$N$ nodes to form a set of random nodes as \(\mathcal{R}\)  ;\\
    & (ii) Each node in \(\mathcal{R}\) discloses one of their connections to form a set of random connections as  \(\mathcal{RN}\);\\ 
    & (iii) Each node in \(\mathcal{RN}\) discloses its degree;\\ 
    & (iv) Select the top \emph{s} highest degree nodes from \(\mathcal{RN}\) as seeds;\\ 
    & (v) Repeat (i)-(iv) for 10,000 times to generate a seed matrix with a dimension of $10,000$\(\times\)$s$.\\ 
 \hline
  \multicolumn{2}{|l|}{3. Diffusion model for a given threshold $\theta$. } \\
 \hline
   & (i) Each node is assigned with a threshold $\theta$ and a starting state value of $0$;\\
   & (ii) Change the state values of nodes to $1$ according to the $k$-th row of the seed matrix, in time step ${t}_0$; for every following time step, check the weighted state value of each node$^,$s connections ($w_i$) and if $w_i\geq \theta$, change the state value of node $i$ to $1$; stop if the number of activated nodes at the current time step $t$ is equal to the previous time step $t-1$; record the final activated number of nodes; \\
   & (iii) Change the $k$ in (ii) from 1 to 10,000 and 
   so that all the combinations in the seed matrix are tried out to kick off the diffusion process. It therefore produces 10,000 possible final activated fractions ${AF}_t$ under this current combination of network, seeding strategy, seed set size $s$ and threshold $\theta$.\\  
 \hline
    \end{tabular}
    \caption{Simulation procedures for a given network, seed set size and threshold. }
    \label{tab2}
\end{table}

\section{Results \label{sec3}}

\subsection{Seeding outcomes for synthetic networks and implications for real-world networks\label{sec3-1}}

\begin{figure}[!h]
    \centering
    \includegraphics[width=0.9\linewidth]{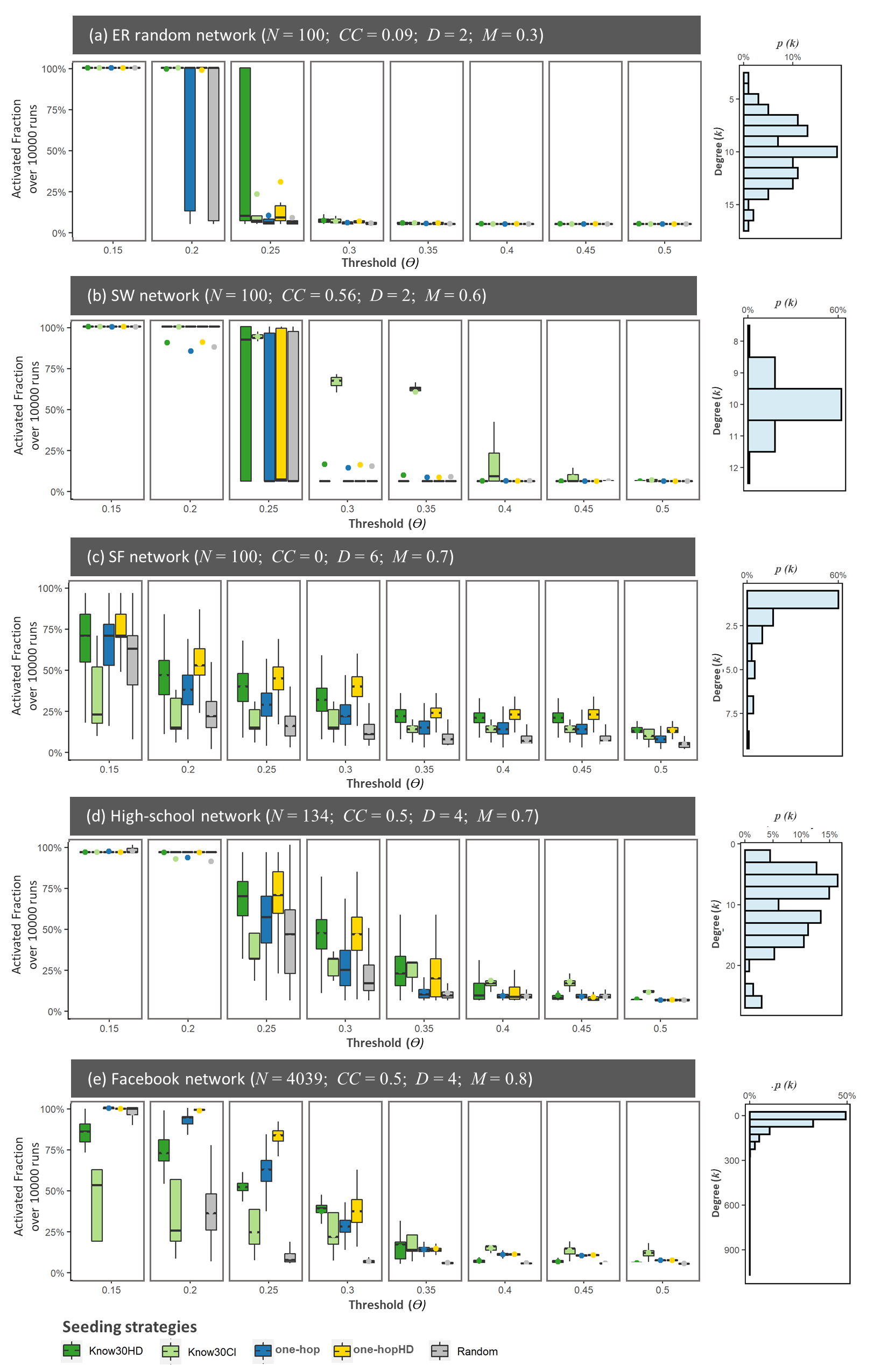}
    \caption{Final activated fractions in (a) ER random network, (b) SW network, (c) SF network, (d) high-school friendship network and (e) Facebook network. The $x$-axis is the threshold of the contagion model, ranging from $0.15$ to $0.5$. For each network, $5\%$ nodes are activated following different seeding strategies. The box-plot shows the $97.5$th, $75$th, $50$th, $25$th and $2.5$th quantiles of the final activated fractions for the $10,000$ trials. Noted are the network attributes, such as size ($N$), clustering coefficient ($CC$), average shortest path length ($D$) and modularity ($M$). Histograms on the right show the degree distribution for the corresponding network.}
    \label{fig2}
\end{figure}

In Fig. \ref{fig2} we display the simulation outcomes for five seeding strategies: one-hop, one-hopHD, Know30HD, Know30Cl and random, for three types of synthetic networks (ER, SW and SF), as well as those for two real-world networks (high school and Facebook; cf. Table \ref{tab1}). The figure shows the activated fractions of agents under thresholds ranging from 0.15 to 0.5. Lower threshold values lead to higher activation percentages, as it must: e.g., an agent with threshold $\theta=0.2$ and with $10$ connections in a network will become activated if two or more connections were activated, in contrast to an agent with $\theta=0.5$ and with $10$ connections requiring at least five activated connections before following suit. The seed set size in Fig. \ref{fig2} is $5\%$, i.e., each seeding strategy can activate $5\%$ of the nodes to kick off the activation diffusion process. As described in Table \ref{tab2}, for every specific threshold and seed set size and any given network, each seeding strategy has been simulated 10,000 times. The variability in the final activated fractions, due to randomness in the activation diffusion process over these 10,000 runs are captured by the error bars in Fig. \ref{fig2}.

Each synthetic network represents a unique network architecture. ER networks describe a theoretical world when human connections are made all by random chances. The networks are characterized by short path lengths but low clustering without community structures. The degree distribution, i.e., the probability distribution $P(k)$ of degree $k$ in the networks, follows a normal distribution. The latter attribute means that the percentage of nodes having degrees much higher or lower than the mean degree $k_{\text{mean}}$ decreases sharply with $|k-k_{\text{mean}}|$. In such (homogeneous) networks, most nodes have similar roles in accelerating or slowing down the activation diffusion process. This explains why the differences between seeding strategies are only noticeable in a narrow band of thresholds. In the example shown in Fig. \ref{fig2}a, differences between seeding strategies are only observed when thresholds are 0.2 and 0.25. For $\theta=0.2$, most seeding strategies except one-hop and random seeding can trigger adoption in the whole network by seeding only $5\%$ nodes, in all the 10,000 random trials. The one-hop and random seeding strategies are the least efficient ones. For both strategies, if they can only seed $5\%$ of the nodes, the chances for their seeds to trigger adoption in the whole network is less than $2.5\%$. When threshold increases to 0.25, the difficulty to pass down diffusion increases. In this case, Know30HD is the only strategy that still has some chance to reach activation fractions that are larger than $50\%$, while other strategies end up with final activation fractions lower than $25\%$ in nearly all the trails. In such a synthetic network wherein every connection is made by random chance, the only significant distinguishing factor among the nodes is their degrees. Know30HD turns out to be the most efficient seeding strategy, followed by one-hopHD. 

SW networks are known for their reproduction of the ``small-world'' properies observed in real social networks, i.e., short path length and high clustering. Compared to ER networks, SW network has community structures with closed triads (cliques) linking nearest nodes and a few shortcuts connecting different communities \cite{watts_collective_1998}. Degree variation in SW networks is even lower than that in ER networks. Rewiring from a regular lattice, most nodes in SW networks used here have exactly the same degree of 10. Only the few nodes rewired to create new connections have degrees that are either 9 or 11, as shown in the degree distribution in Fig.\ref{fig2}b, which approximates a delta distribution (each
node has a degree of exactly $k$). As degree differences between nodes in SW networks are not present, Know30HD is not longer the most efficient strategy. Instead, Know30Cl stands out as the most efficient seeding strategy even for high thresholds. This is as expected since clustered seeding is designed for complex contagion under the small-world nature \cite{Centola+2018, 10.1086/521848}. Under the standardized community structure in SW networks, seeding a cluster of nodes is essential to overcome the thresholds of nodes connected to the same cluster \cite{Centola+2018}. Moreover, since there are clustered structures present in SW networks, connections among different clusters close to each other are plentiful \cite{watts_collective_1998,Centola+2018}. Such connections create channels for contagion passing from one cluster to another, and consequently, throughout the entire network. In such a network with closely knitted community structure and low degree variances, one-hopHD strategy is not efficient.

SF networks mimic high degree variances observed often in real-world networks. The degree distribution of SF networks follows a power-law (see right panel of Fig. \ref{fig2}(c)). It results in a hub-dominated network architecture: hubs are those (few) nodes that are highly connected while most nodes having degrees that are much less than the mean.  In such a synthetic network dominated by hubs, one-hopHD stands out as the most efficient seeding strategy, as shown in Fig. 2c. The strategy used by one-hopHD, i.e., nomination by random nodes and then high-degree seeding, turns out to be more efficient than purely high-degree seeding by Know30HD, though the later has access to more degree information. Clustered seeding is no longer applicable in such a network structure, which ends up with final activated fractions that are even lower than one-hop and random seeding.

These results in the synthetic networks demonstrate that seeding outcomes of the five strategies depend strongly on the architecture of the concerned network: while community structure (or high clustering) makes clustered seeding the most suitable strategy, when high degree variance is taken into account, one-hopHD is the most efficient one. However, real-world social networks are neither pure SW nor pure SF networks but a mixture of both high degree variances and community structures. For a comparison with the theoretical results from the synthetic networks, we therefore use two further real-world networks. 

The first one is a high school friendship network representing a small-size ($N=134$) social network that is very similar to the synthetic SW networks in terms of high clustering and low path length (Fig. \ref{fig2}(d)). Nevertheless, the approximately delta degree distribution for SW networks does not hold for this network, as shown as the log-normal degree distribution in the right panel of Fig. \ref{fig2}(d). Under such a mixture of high clustering and moderate degree variances, Know30HD and one-hopHD are the two most effective strategies. Both of them provide higher final activated fractions than Know30Cl in most cases. When threshold goes up to 0.35, Know30Cl outperforms these two but the margin is relatively small. The second network is a large-size ($N=4039$) Facebook network. It has the same clustering coefficient ($CC=0.5$) and average shortest path lengths ($D=4$) as the high school network, but its degree distribution follows a power law. In other words, it inherits the extremely high degree variances from the SF network, while maintaining high clustering and low path lengths that is similar to SW network. With high variations in node degrees, one-hopHD stands out as the most effective seeding option, leading to final an activated fraction that is much higher than those of clustered seeding (Know30Cl) and purely high-degree seeding (Know30HD). It seems that under the co-existence of high degree variances, high clustering and short path length, one-hopHD is a promising strategy and its performance stands out even more when the degree variance increases.

In the following subsections, we further validate the above observations for more real-world networks.

\subsection{Comparison between the outcomes of one-hopHD and Know30HD strategies for real-world networks\label{sec3-2}}

In this subsection, we compare the seeding outcomes by one-hopHD and Know30HD in ten real-world networks and analyze if the combination of searching from random connections and then high-degree seeding is more efficient than purely high-degree seeding.

\begin{figure}[!h]
    \centering
    \includegraphics[width=1\linewidth]{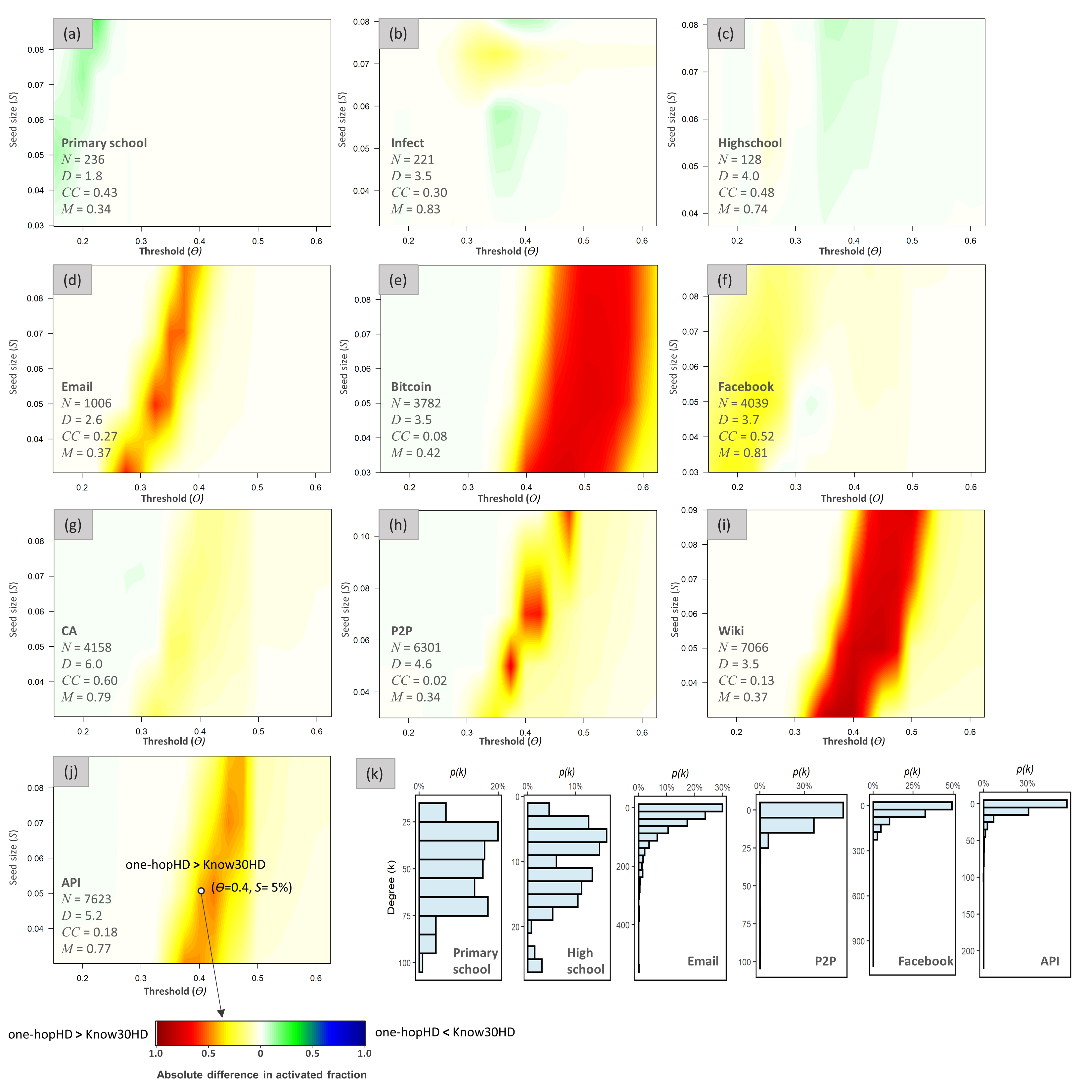}
    \caption{Comparison of the final activated fractions between one-hopHD and Know30HD in 10 real-world networks (a-j) and degree distributions of selected networks (k).The \textit{x}-axis is the value of the threshold used in the contagion model, and the \textit{y}-axis is the seed set size. The colour gradient in (a-j) represents the absolute difference in the final activated fractions between one-hopHD and Know30HD (averaged of the 10,000 runs), under specific combinations of threshold and seed set size. The coordinates of every color pixel correspond to a specific threshold (value of \textit{x}-axis) and seed set size (value of \textit{y}-axis), respectively. Colours of red and yellow indicate one-hopHD leads to higher activated fractions than Know30HD (one-hopHD$>$Know30HD), and the opposite for blue and green colours. For example, the highlighted point in (j) shows that the final activated fraction achieved by one-hopHD is larger than that of Know30HD in the API network when the threshold is 0.4 and seed set size is 5$\%$. (k) Degree distributions for selected networks, in which the vertical axis is the degree ($k$) and horizontal axis is the probability of degree ($p(k)$). Degree distributions of the other networks can be found in Supporting Information Fig. S1. Compared to small-size offline networks (e.g., primary and high school networks), large online social networks (e.g., Email, P2P, Facebook and API) have much higher degree variances following a power-law decay. }
    \label{fig3}
\end{figure}

The differences between one-hopHD and Know30HD seeding outcomes in ten networks are shown in Fig. \ref{fig3}(a-j), for a wide range of threshold values and seed set sizes. The $x$-axis is the threshold value, ranging from 0.15 to 0.6. The $y$-axis is the seed set size within the range of 3$\%$ to 10$\%$. The color gradient is the absolute difference between the final activated fractions (again, averaged over 10,000 runs) achieved by one-hopHD and Know30HD, in which gradients from yellow to red indicates that one-hopHD leads to higher activated fraction than Know30HD and vice versa for gradients from green to blue. Each panel in Fig. \ref{fig3}(a-j) corresponds to the results in a specific network, under every possible combination for seed set size and threshold. The coordinate of every color pixel within each panel corresponds to the specific seed set size and threshold value. The highlighted point in Fig. \ref{fig3}(j), for example, shows that the final activated fraction achieved by one-hopHD is larger than that of Know30HD in the API network when $\theta=0.4$ and 5$\%$ nodes are used as seeds.

The networks used for comparison here cover offline and online social networks with various sizes and characteristics. The first three networks in Fig. \ref{fig3}(a-c) are offline social networks of a primary school, a high school and an exhibition, respectively. Each of them includes less than 250 nodes and has the common characteristics of a small-world network as being highly clustered and having short path length. The degree distribution follows a log-normal distribution (e.g., first two subpanels in Fig. \ref{fig3}(k)), indicating moderate degree variances. The observations we made in Fig. \ref{fig2}(d) hold true for these networks with small-world like characteristics and moderate degree variances. For example, recall that Fig. \ref{fig2}(d) corresponds to a specific case of the high school network, with seed set size of $5\%$, for which one-hopHD and Know30HD are more effective than the other seeding options. Among these two, Know30HD leads to slightly better seeding outcomes than one-hopHD, as shown in the green areas in Fig. \ref{fig3}(a-c). In other words, for small-world like networks with moderate degree variances, the proposed one-hopHD strategy is not necessarily the best way of seeding. Its seeding outcomes can get easily caught up by purely high-degree seeding as long as the later has access to the degree information of more nodes.

The other seven networks from Fig. \ref{fig3}(d-j) are online social networks for email communication, Bitcoin trading, Facebook friendship, academic author collaboration (CA), p2p, Wiki voting and API, respectively, including from 1,000 to 7,600 nodes. Compared to the offline networks, such online networks have similar or even shorter path lengths and much more skewed degree distributions following a power-lay decay (e.g., last four subpanels of Fig. \ref{fig3}(k)). In these online networks, one-hopHD outperforms Know30HD in nearly every combination of seed set size and threshold considered here (Fig. \ref{fig3}(d-j)), even though Know30HD has the degree information of more nodes. In some cases, the absolute difference is as large as 1 (red areas in Fig. \ref{fig3}(d-j)), indicating that one-hopHD can lead to adoption in the whole network while Know30HD fails to trigger any significant diffusion. It suggests that, for such large networks with high degree variance and good connectivity (i.e., relatively short path lengths), one-hopHD is much more efficient than Know30HD and the premium cannot be caught up by purely high-degree seeding even though the later one has access to more degree information. 

Moreover, the efficiency of one-hopHD in (Fig. \ref{fig3}(d-j)) is independent of the clustering coefficients. The networks in (Fig. \ref{fig3}(d-j)) have clustering coefficients varying from the lowest of 0.08 in the Bitcoin network (Fig. \ref{fig3}(e)) to the highest of 0.6 in the CA network (Fig. \ref{fig3}(h)). The dominant factors deciding the relative performance between one-hopHD and Know30HD seem to be the degree variance and connectivity of the network. As demonstrated in previous studies \cite{Crucitti2004ErrorAA,Tsiotas6701}, networks with high degree variances such as those with a power-law decay end up in a system that most nodes have very few links while a few nodes (or hubs) have unproportionally large numbers of links. These few hubs can reach the majority of the network within one step. In turn, searching the connections from a random node is very likely to reach these hubs \cite{Crucitti2004ErrorAA}. As a result, the tactic used by one-hopHD, i.e., searching from random connections before high-degree seeding, is very efficient to reach the highly connected nodes. This is the key why one-hopHD is efficient in such inhomogeneous networks with high degree variances. 

\subsection{Comparison between the outcomes of one-hopHD and Know30Cl for real-world networks\label{sec3-3}}

An important seeding strategy for complex contagion is clustered seeding: seeding a cluster of nodes that are connected to each other so that the other nodes connected to this cluster can receive multiple exposures to overcome their thresholds \cite{Centola+2018, 10.1086/521848}. In order to evaluate the effectiveness of one-hopHD under complex contagion, its seeding outcomes should therefore be compared with clustered seeding with limited network information. Specifically, clustered seeding strategy (Know30Cl) assumes that the degrees of $30\%$ nodes are known (``known nodes''), and seeds are the highest degree node chosen from these known nodes, and its $s-1$ direct connections. In contrast, one-hopHD picks the highest degree nodes out from the random connections and has no guarantee that the seeds are direct connections of each other. According to the justification for clustered seeding, one-hopHD would be suboptimal since seeds are scattered throughout the network and do not provide the required multiple exposures to other nodes adjacent to the seeds. 
\begin{figure}[!h]
    \centering
    \includegraphics[width=1\linewidth]{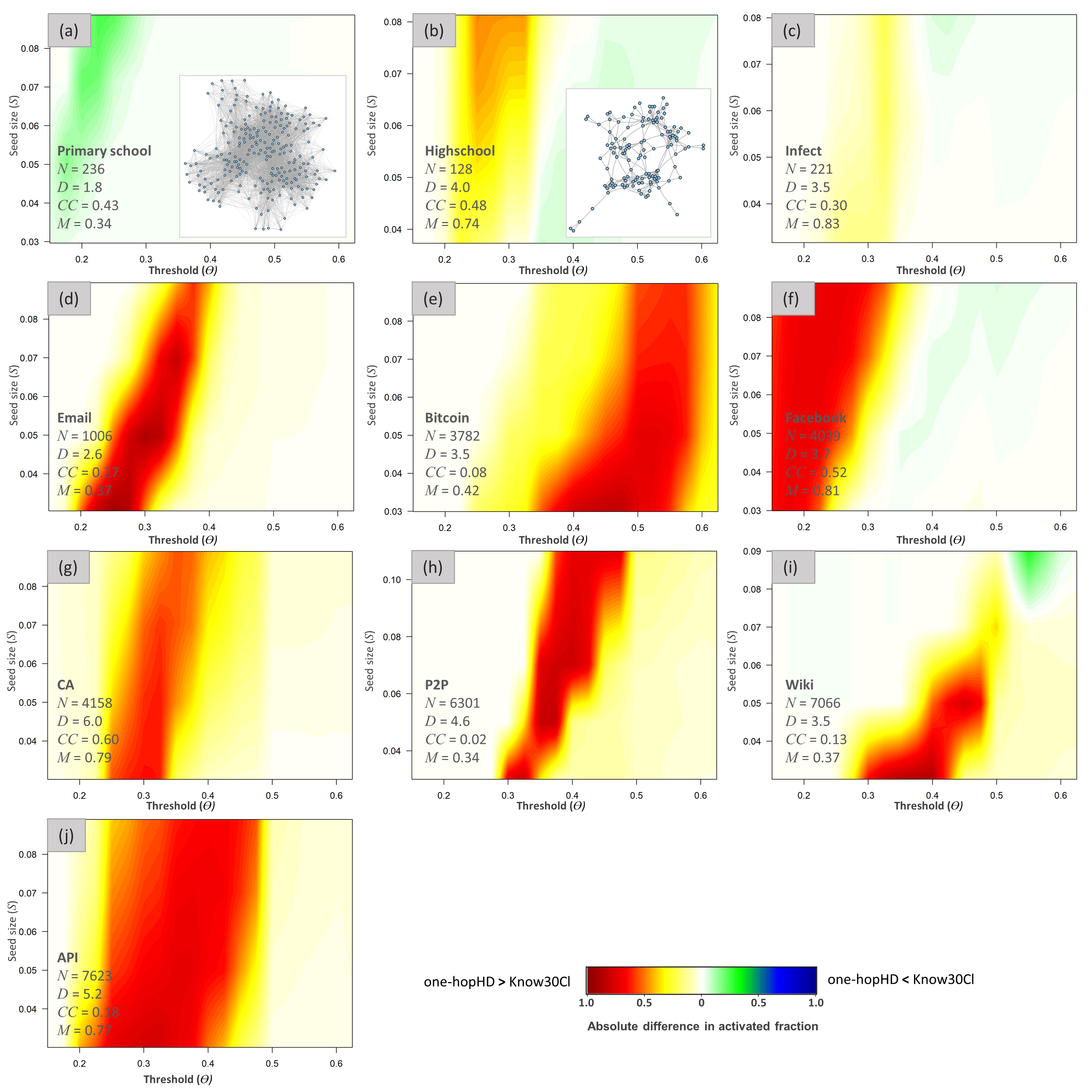}
    \caption{Comparison of the final activated fractions between one-hopHD and Know30Cl in 10 real networks (a-j).The \textit{x}-axis is the value of the threshold used in the contagion model, and the \textit{y}-axis is the seed size. The colour gradient in (a-j) represents the absolute difference in the final activated fractions between one-hopHD and Know30Cl (average of the 10,000 runs), under specific combination of threshold and seed set size. The coordinates of every color pixel correspond to the threshold (value of \textit{x}-axis) and seed set size (value of \textit{y}-axis), respectively. Colours of red and yellow indicate one-hopHD lead to higher activated fractions than Know30Cl (one-hopHD$\>$Know30Cl), and the opposite for colours of blue and green. The primary school and high school networks are visualized in a$\&$b, in which the network topology of the high school is less homogeneous.}
    \label{fig4}
\end{figure}

Simulation involving our ten real-world networks, however, suggests that one-hopHD outperforms the clustered seeding strategy (Know30Cl) in most cases, as shown in Fig. \ref{fig4}(a,d-j), where we compare the final activated fractions by one-hopHD to those of Know30Cl. Specifically, with the exception of the primary school network (Fig. \ref{fig4}(a)) and high-threshold scenarios in a few networks, one-hopHD provides much higher final activated fractions than Know30Cl. This is especially true for large networks (Fig. \ref{fig4}(d-j)), where the absolute difference between the final activated fraction of one-hopHD and Know30Cl is as high as unity. These results indicate that, by activating the same number of seeds, one-hopHD successfully diffuses the contagion to the whole network while seeds of Know30Cl fail to activate substantial amounts of nodes, calling into question the justification for clustered seeding.

The reason why Know30Cl leads to worse seeding outcomes may lie in the fact that its effectiveness is mostly demonstrated in homogeneous networks such as synthetic SW networks (i.e., most nodes have similar degrees and there are abundant connections between every two clusters, meaning that the networks have low modularity values) \cite{Centola+2018, 10.1086/521848}. A homogeneous network ensures that seeding a cluster can activate the nodes adjacent to the seeds, and play the crucial role of passing the contagion on from one cluster to another by the redundant connections among clusters. This reasoning is bolstered by the observation that among all the real networks used for simulations in Fig. \ref{fig4}, the primary school network is the most homogeneous, and also the only network for which Know30Cl performs better than one-hopHD. Most nodes in the primary school network have degrees around the mean value of 50, which, as shown in Fig. \ref{fig4}(a), results in a closely and evenly knitted network. Seeding a cluster of nodes in such a homogeneous network triggers large scale diffusion as predicted by \cite{Centola+2018, 10.1086/521848}. 
\begin{figure}[!h]
    \centering
    \includegraphics[width=0.8\linewidth]{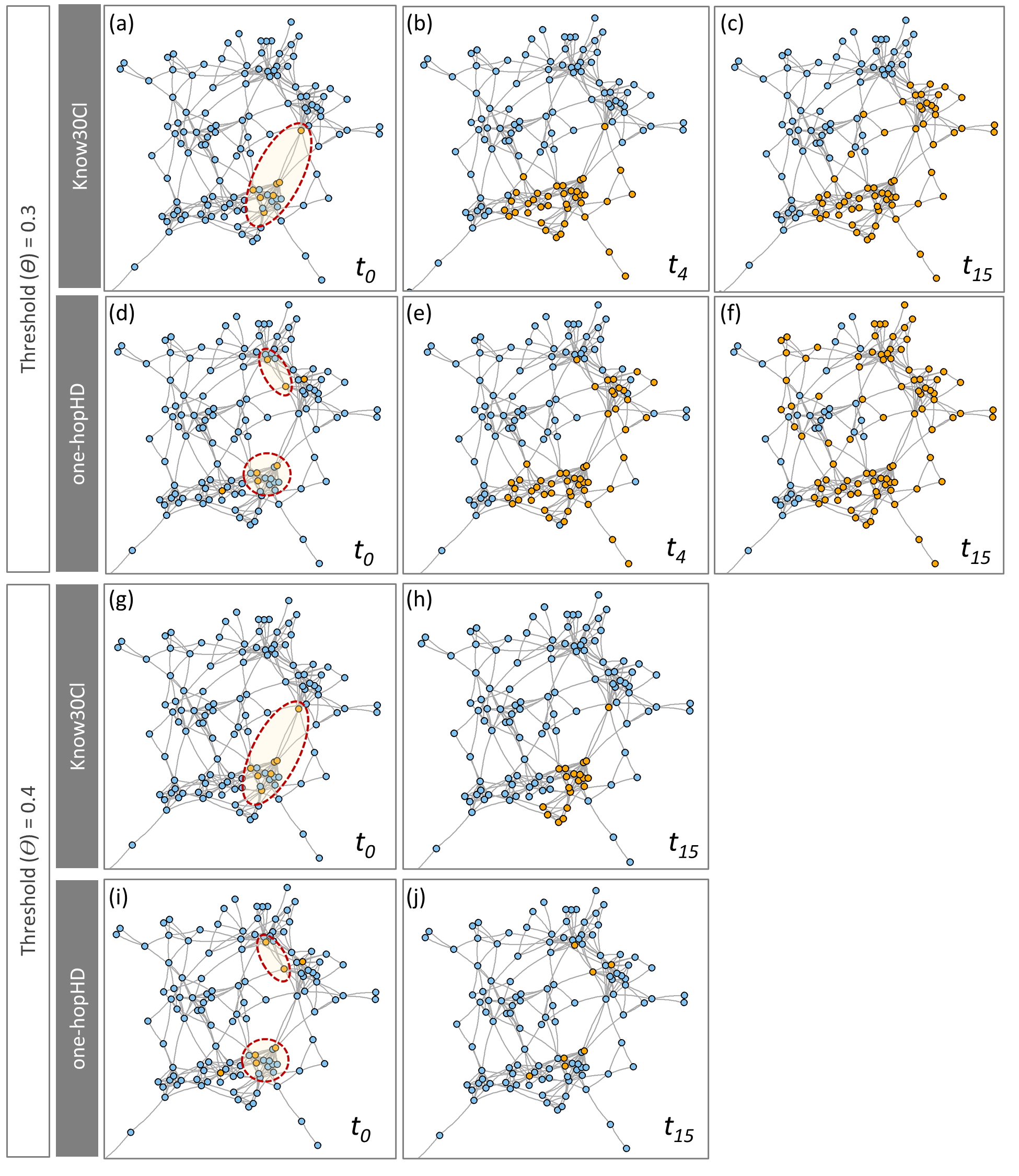}
    \caption{Diffusion process in the high school network driven by the seeds of Know30Cl and one-hopHD. Nodes in yellow indicate that they are activated, while inactivated nodes are in blue. At time $t_0$, Know30Cl always activates a cluster of seeds, which are direct connections of each other as highlighted in the red circle in (a) and (g). On contrast, the one-hopHD starts from partly connected nodes, as highlighted in two red circles in (d) and (i). When the threshold is not too high (e.g., being 0.3 in this example), the highly clustered seeds provided by Know30Cl turn out to be overly concentrated in one area and fail to reach other parts of the network, as shown by the diffusion process from (a) to (c). Seeds of one-hopHD, on the contrary, can reach the majority of the network (f) since they start from more than one clusters. When the threshold is increased to 0.4, seeds need to be highly clustered to activate any cascade in the network. Know30Cl can activate slightly more nodes that one-hopHD, referring to the panels of (h) and (j). But the margin provided by Know30Cl is very low since even clustered seeding would fail to activate significant percentage of nodes when threshold is high in an inhomogeneous network.}
    \label{fig5}
\end{figure}

For other real-world networks, specifically with less homogeneous architecture, clustered seeding may not work well since the seeds in a clustered seeding strategy are overly concentrated in the same cluster, combined with a situation that there may not be enough connections between clusters to pass the contagion on. Let us demonstrate this concept using the high school network as an example. As for network architecture, the high school network has a similar size and clustering coefficient but a less homogeneous architecture in comparison to the primary school network, highlighted by a more skewed degree distribution and a higher value of modularity. Moreover, the modularity value in the high school network doubles that of the primary school network.

The seeding outcomes for Know30Cl and one-hopHD strategies for the high school network are shown in Fig. \ref{fig5}. At time $t_0$, the two strategies activate the same number of seeds ($5\%$ of the nodes). The seeds selected by Know30Cl are always direct connections of each other, as highlighted in the red circle in Fig. \ref{fig5}(a). In contrast, seeds selected by one-hopHD are not necessarily directly connected, but are sometimes partly connected as shown in Fig. \ref{fig5}(d). When the threshold is not too high (such as being $0.3$), the partly connected seeds provided by one-hopHD turn out to be efficient enough to activate 72$\%$ of nodes in the network, as shown in Fig. \ref{fig5}(d-f). Under the same threshold value, seeds of Know30Cl are a group of directly connected nodes. Their activation reaches and activates their common connections within the close vicinity fast, within the first few time steps (Fig. \ref{fig5}(a-b)). As the diffusion process goes on, however, they fail to reach the other parts of the network where the connections are sparse (Fig. \ref{fig5}(c)). The clustering of seeds becomes necessary again only when the threshed is high. For example, when the threshold goes up to 0.4 in Fig. \ref{fig5}(g-j), the partly connected seeds provided by one-hopHD fail to overcome the thresholds of any nearby nodes (Fig. \ref{fig5}(i-j)), while seeds of Know30Cl can still activate the nodes directly connected to the seeds (Fig. \ref{fig5}(g-h)). The difference between two strategies is however quite limited when the threshold is high, as shown in Fig. \ref{fig5}(h). 

The above example shows that the key for the success of one-hopHD under complex contagion is to have seeds that are partly connected. Though one-hopHD does not deliberately use clustering as a criteria of seed selection, the selected seeds are often partly connected due to the connectedness of high degree nodes in most networks. For the ten real-world networks analyzed here, despite large variations of network sizes ranging from 100 to 7,600, the average shortest path lengths between the top 10$\%$ highest degree nodes are between 1.3 to 4.5 (Table \ref{tab1}). The academic author collaboration network (CA) has the highest value of 4.5, for a network size of $N=4158$. For this network with the highest path length between highest degree nodes, one-hopHD still leads to higher final activated fractions than Know30Cl, though its relative efficiency (Fig. \ref{fig4}(g)) is less significant than what we observed in other networks (Fig. \ref{fig4}(d-j)).

\subsection{Comparison between the outcomes of one-hopHD, one-hop and random seeding for real-world networks\label{sec3-4}}

\begin{figure}[!h]
    \centering
    \includegraphics[width=1\linewidth]{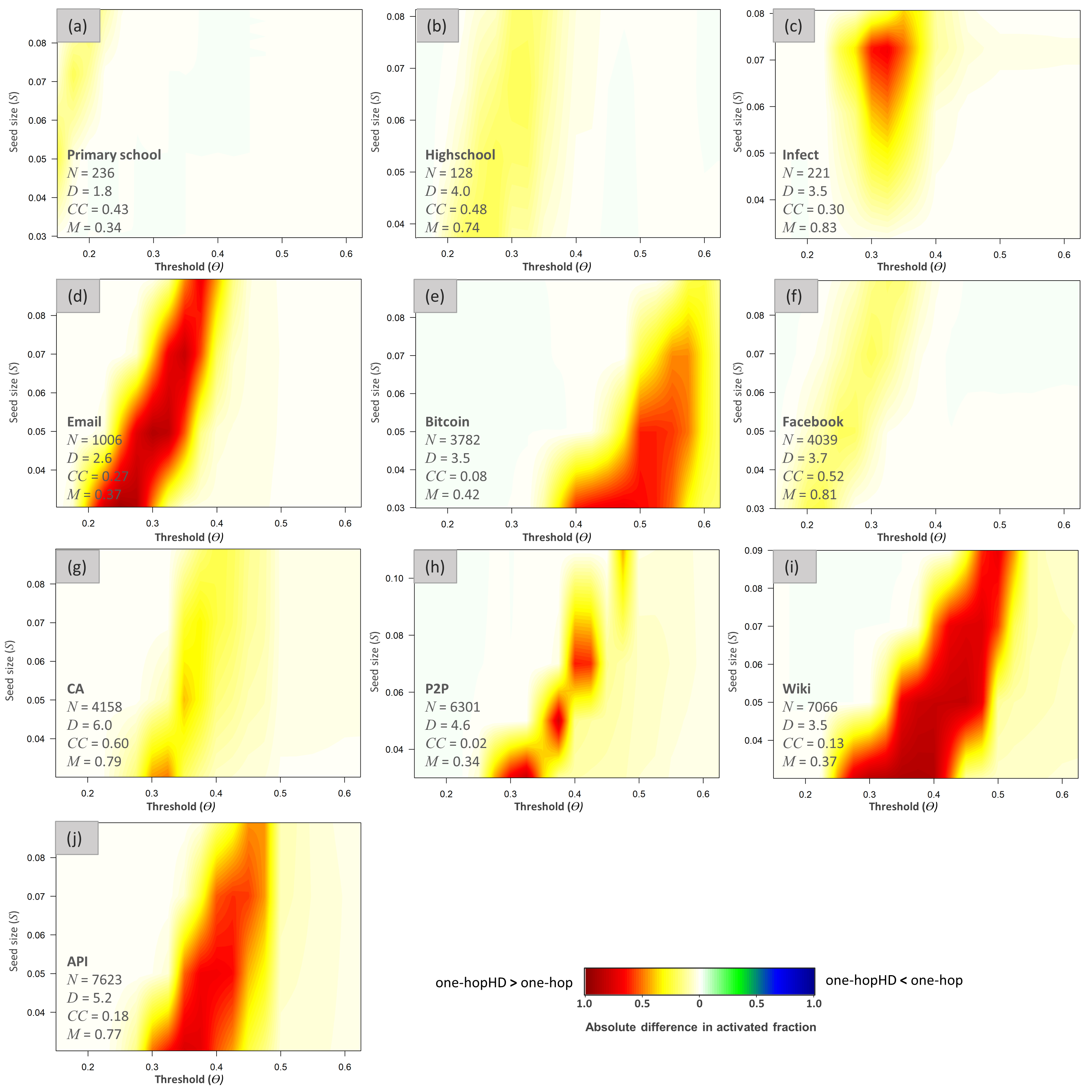}
    \caption{Comparison of the final activated fractions between one-hopHD and one-hop in 10 real networks (a-j), with the same number of seeds. The $x$-axis is the value of threshold used in the contagion model, and the $y$-axis is the seed size. The colour gradient in (a-j) represents the absolute difference in the final activated fractions between one-hopHD and one-hop (average of 10,000 runs), under specific combination of threshold and seed set size. The coordinates of every color pixel corresponds to threshold (value of $x$-axis) and seed set size (value of $x$-axis), respectively. With the same number of seeds, one-hopHD leads to higher final activated fractions than one-hop in all the networks tested here.}
    \label{fig6}
\end{figure}

Compared to one-hopHD, both the conventional one-hop and random seeding require less network information. If all these strategies are used to activate the same number of seeds, as expected, one-hopHD would lead to better seeding outcomes. Indeed, as shown in Fig. \ref{fig6}, one-hopHD leads to higher activated fraction than one-hop strategy in all the networks tested here, under various settings of threshold and seed set size. Results for the comparison between one-hopHD and random seeding are similar, which can be found in the Supporting Information Fig. S2.
\begin{figure}[!h]
    \centering
    \includegraphics[width=1\linewidth]{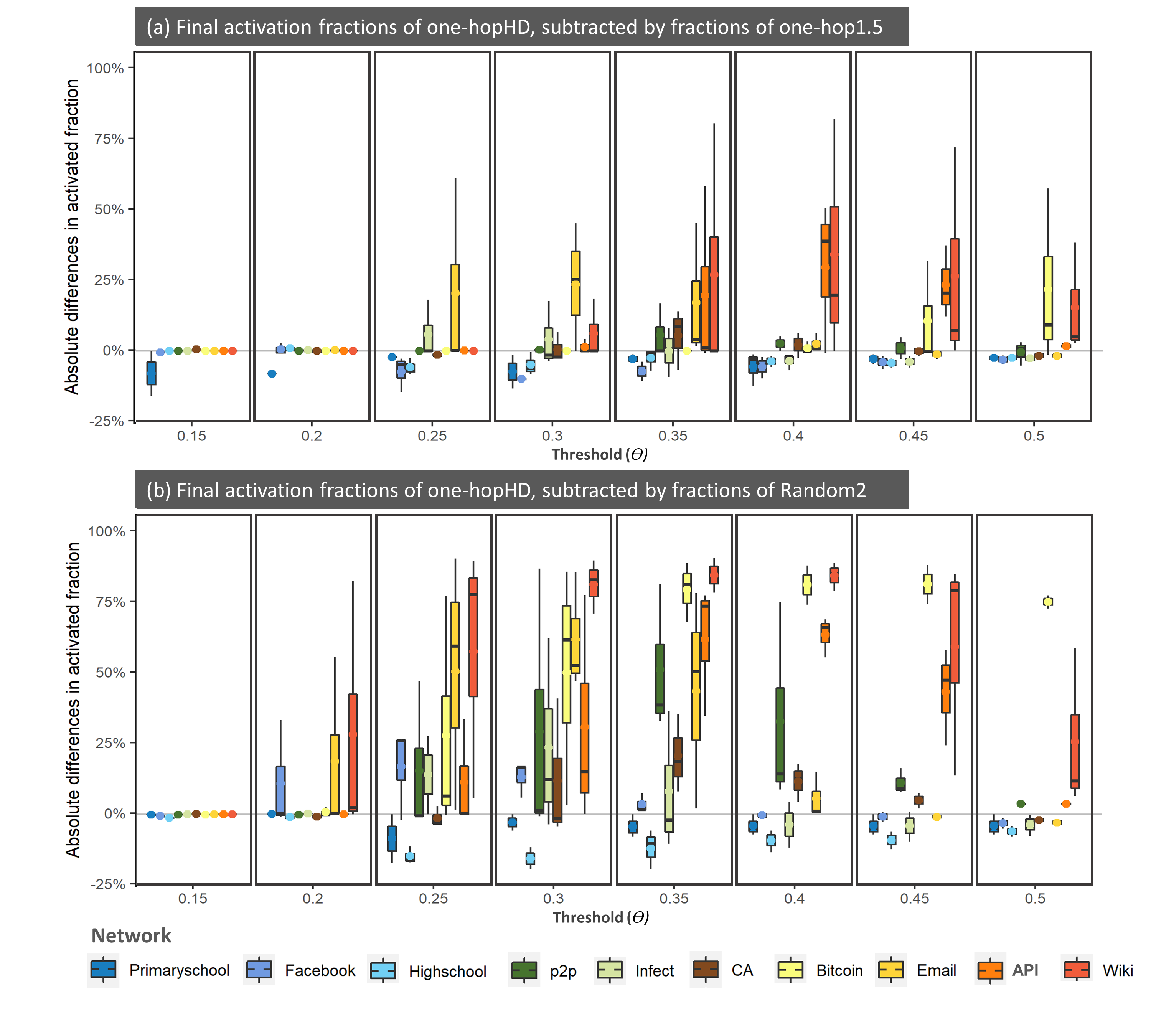}
    \caption{Comparison between one-hopHD strategy with one-hop and random seeding with increase seed set size (i.e., one-hop1.5 and Random2). (a) The final activated fractions of one-hopHD with seed sizes of 3, 5 and $7\%$ were subtracted by the fractions achieved by one-hop1.5 with seed sizes of 4.5, 7.5 and 10.5$\%$, respectively, for the same network and a given threshold. (b) The final activated fractions of one-hopHD with seed sizes of 3, 5 and 7$\%$ were subtracted by the fractions achieved by Random2 with seed sizes of 6, 10 and 14$\%$, respectively. Every box displays the minima, 25th percentile, median, 75th percentile and the maxima of the differences. The relative efficiency of one-hopHD largely remains even when the seed size of one-hop and random seeding increases by $50\%$ and $100\%$, respectively.}
    \label{fig7}
\end{figure}

One critique of the one-hop strategy is the generality of its efficacy, which varies a lot from one network to another, depending on the architecture of the network and the setting of contagion models such as seed set size and threshold value \cite{Kumar2019,Chin2021}. Such variabilities in the performance are to be expected, given that the seeding strategy is applied stochastically, starting from a random draw of nodes from the network. Seeds provided by one-hop are characterized by the use of high-degree nodes (normally higher than the mean degree). They are also likely to be distributed broadly throughout the network, though to a lesser extent than for the case when seeding is random. The one-hopHD stresses the first characteristic even more by selecting the high-degree nodes from random connections. In turn, the second characteristic is compromised as seeds would be more concentrated towards the ``center'' of the network. The center here refers to the innermost core of the network where most efficient spreaders locate \cite{Kitsak2010}. Such a trade-off between high degree and randomness turns out to be successful in increasing the final activated fraction in a wide range of network structures, seed set sizes and threshold value (Fig. \ref{fig6}). Compared to the one-hop strategy, one-hopHD provides better seeding outcomes across different networks, seed set sizes and thresholds. 

Recognizing that one-hop and random seeding require less network information than one-hopHD, we test whether the premium yielded by one-hopHD can be easily caught up by one-hop or random if the latter can seed more. Specifically, the number of seeds by one-hop and random are increased by $50\%$ (``one-hop1.5'') and $100\%$ (``Random2''), respectively. Results are shown in Fig. \ref{fig7}. In Fig. \ref{fig7}(a), the final activated fractions achieved by one-hop by seeding 4.5, 7.5 and 10.5$\%$ nodes were compared to the results achieved by one-hopHD by seeding 3, 5 and 7$\%$ nodes, respectively. Each box plot shows the variations of differences between two seeding strategies for the same network and a given threshold. For example, the first box in Fig. \ref{fig7}(a) shows the differences between one-hopHD and one-hop1.5 for the primary school network when the threshold is 0.15. When one-hop can seed 50$\%$ more nodes, its final activated fraction can indeed catch up with one-hopHD in a few networks such as the primary school network, Facebook network and high school network. For the other networks, the final activated fractions by one-hop1.5 still lag behind one-hopHD, especially for the networks of Bitcoin, Email, API and Wiki. In short, while one-hop can catch up the seeding outcomes of one-hopHD in some cases simply by increasing the number of seeds, the number of seeds need to be increased by at least $50\%$ and it does not apply to every network. In fact, for half the networks tested here, a $50\%$ increase of the seed number of one-hop still fails to reproduce the seeding outcomes achieved by one-hopHD.

Regarding random seeding that requires no network information at all, its number of seeds are doubled and compared to one-hopHD. In Fig. \ref{fig7}(b), random strategy seeds 6, 10 and 14$\%$ nodes in each network and the final activated fractions were compared to those achieved by one-hopHD which can only seed 3, 5, and 7$\%$ nodes in the same network. Even though random seeding doubles the number of seeds, its seeding outcomes still largely falls behind those of one-hopHD in most networks. 
The absolute differences between one-hopHD and Random2 are as high as $50\%$ to $80\%$ in half of the cases. The premium yielded by one-hopHD in terms of the final activated fraction largely remains even though random strategy can seed twice as many. 

The above results suggest that the conclusion found for simple contagion \cite{Akbarpour2017}, i.e., a small increase of the random seed set size can match the seeding outcomes with network information, is not applicable to complex contagion. Our results here suggest that network information, even if it is very limited, {\it does\/} matter when the diffusion process is governed by complex contagion. With the help of some degree and connection information, seeding strategies such as one-hopHD can lead to significantly higher final activated fractions that cannot be matched by blindly seeding more.

\section{Conclusion\label{sec4}}

This study has introduced the one-hopHD seeding strategy, a variant of the conventional one-hop strategy, for effectively seeding contagions in networks with very limited information on nodes and edges. The conventional one-hop strategy selects {\it high-degree\/} nodes in a {\it random\/} fashion. The first characteristic attempts to select potentially influential nodes and the second one hopes to have the seeds distributed broadly throughout the network. The one-hopHD strategy proposed here enhances the first characteristic by selecting the high-degree nodes from the random connections provided by the conventional one-hop strategy. While such a strategy compromises the randomness of conventional one-hop seeds, it reaches a better balance between reaching influential nodes and scattering seeds broadly. It outperforms alternative seeding strategies from the literature across a wide range of networks and parameter settings. 
\begin{table}[H]
\resizebox{\textwidth}{!}{%
    \centering
    \begin{tabular}{c|c|c|c|l}
        \hline
        \textbf{Networks} & \textbf{Path lengths} &\textbf{Clustering} & \textbf{Degree Variance} & \hfill\textbf{Seeding strategies comparison}\hfill \\
        \hline
         Synthetic ER network & Short & Zero & \begin{tabular}{@{}c@{}} Moderate \\ (normal distribution)\end{tabular}  & \begin{tabular}{@{}l@{}} Know30HD $>$ one-hopHD $>$ Know30Cl $>$  \\ one-hop $>$ random\end{tabular} \\
         \hline
        Synthetic SW network & Short & High  & \begin{tabular}{@{}c@{}} Nearly zero \\ (approx. delta distribution)\end{tabular}  & \begin{tabular}{@{}l@{}} Know30Cl $>$ one-hopHD, Know30HD $>$ \\ one-hop $>$ random\end{tabular} \\
        \hline
        Synthetic SF network & Short & Zero & \begin{tabular}{@{}c@{}} High \\ (power law distribution)\end{tabular} & \begin{tabular}{@{}l@{}} one-hopHD $>$ Know30HD $>$ one-hop $>$ \\ Know30Cl $>$ random\end{tabular} \\
        \hline
         \begin{tabular}{@{}l@{}} Small offline networks: Primary school, \\ high-school and exhibition networks \end{tabular}  & Short & High & \begin{tabular}{@{}c@{}} Moderate \\ (log-normal distribution)\end{tabular}  & \begin{tabular}{@{}l@{}} Know30HD $>$ one-hopHD, Know30Cl $>$ \\ one-hop $>$ random\end{tabular} \\
        \hline
        \begin{tabular}{@{}l@{}} Large online networks: Email, Bitcoin,  \\Facebook, CA, P2P, Wiki and API \end{tabular} & Short & Low to high & \begin{tabular}{@{}c@{}} High \\ (power law distribution)\end{tabular}  & \begin{tabular}{@{}l@{}} one-hopHD $>$ one-hop $>$ know30HD $>$ \\ Know30Cl $>$ random\end{tabular} \\   
        \hline
    \end{tabular}}
    \caption{Summary of network properties and seeding outcomes.} \label{tab3}
\end{table}

The comparative performance of the one-hopHD in various networks is summarized in Table \ref{tab3}. The efficiency of one-hopHD especially stands out in networks with high degree variances and short path lengths --- these are networks that are inhomogeneous as well as closely-knit. High degree variance in a network has two implications. The first one is that there exist some highly connected nodes (e.g., hubs) within the network. The second one is that most nodes have very few links, but are connected to the hubs. For such networks, the connections nominated by random nodes (i.e., the first step of one-hopHD) are more often highly connected nodes than the random nodes themselves. This chance is much higher when the extra step of explicitly screening for high degree nodes among nominated neighbors is taken, as in the one-hopHD strategy. Therefore, for networks with high degree variance, the combination of searching from random connections and high-degree seeding is much more efficient than purely high-degree seeding of random nodes. We have demonstrated this by comparing the results of one-hopHD to Know30HD in both the synthetic and real networks. Even though the Know30HD strategy has access to more degree information, i.e., $30\%$ nodes compared to $15\%$ for one-hopHD, its seeding outcomes largely lag behind those of one-hopHD in networks with high degree variances, such as the synthetic scale-free networks or large online social networks with degree distribution approximating a power-law decay. Such close-knit networks also make seeds selected by one-hopHD often partly connected. Combined with the random nature inherent in the random connection nomination process, seeds for one-hopHD are partly connected but are not overly concentrated. These are desirable properties for overcoming thresholds in complex contagion processes, as well as for wide diffusion of activation to different parts of the network when the architecture of the network is not homogeneous. This is in fact the reason why one-hopHD outperforms the clustering seeding strategy with more degree information (Know30Cl) for a wide range of social networks. 

While it is true that one-hopHD requires more network information than the conventional one-hop strategy, in most practical situations the extra information can be secured through reasonable efforts. Importantly, with the help of this extra information, one-hopHD can lead to similar or much better seeding outcomes than the conventional one-hop strategy even though the latter has the option to seed much more. This implies that improvements achieved by the one-hopHD seeding strategy cannot be matched by simply having more seeds. This also holds true when we compare the seeding outcomes of one-hopHD to random seeding even if the latter can seed twice as many. Our results highlight that limited network information to inform seeding under complex contagion and demonstrate that one-hopHD strategy is an efficient way to ensure of limited information on the involved network.

Another limitation of this study is that the efficiency of one-hopHD is not compared to the best possible outcomes achieved by influence-maximization algorithms under full network information. The calculation of such theoretical optima is complex and approximation cumbersome, but a potential direction for future research. We therefore do not know how far below the upper bound on performance we still are and whether yet other strategies could greatly improve on one-hopHD. These limitations not withstanding, the present paper suggests the possibility of greater efficacy of network interventions in applications ranging from public health to the diffusion of sustainable behaviors.

\section*{Acknowledgement}

\textbf{Funding:} This work is part of the project `ENgaging Residents in Green energy Investments through Social networks, complExity, and Design' (ENRGISED), which has been funded by the Netherlands Organization for Scientific Research (NWO).

\noindent\textbf{Author contributions:} 
All the authors conceived the mathematical principles. JO performed the numerical simulation and application to the synthetic and real networks, with help from DP. JO wrote the first draft of the manuscript. All authors reviewed the final text.

\noindent\textbf{Competing interests:} The authors declare no competing interests.

\noindent\textbf{Data and code availability} Network data used in this study can be found in the open databases of SocioPattern (\url{http://www.sociopatterns.org}) and the Standford Large Network Collection (\url{https://snap.stanford.edu/data/#communities}). 
All the code of contagion models, simulation of seeding strategies and estimation of the final activated fractions will be made available online in the repository of Github upon publication: \url{https://github.com/JiaminOu/High-degree-seeding-of-random-connections-in-unknown-graph}. 

\section*{Supplementary Materials}
SI- Supplementary plots on (Fig. S1) Degree distribution of all the social networks used in this study, and (Fig. S2) Comparison of the final activated fractions between one-hopHD and random seeding in 10 real networks, with same number of seeds.

\FloatBarrier

\bibliographystyle{unsrt}

\begin{thebibliography}{10}

\bibitem{Vosoughi1146}
Soroush Vosoughi, Deb Roy, and Sinan Aral.
\newblock The spread of true and false news online.
\newblock {\em Science}, 359(6380):1146--1151, 2018.

\bibitem{doi:10.1056/NEJMsa066082}
Nicholas~A. Christakis and James~H. Fowler.
\newblock The spread of obesity in a large social network over 32 years.
\newblock {\em New England Journal of Medicine}, 357(4):370--379, 2007.
\newblock PMID: 17652652.

\bibitem{doi:10.1056/NEJMoa1003176}
Jennifer~L. Gardy, James~C. Johnston, Shannan J.~Ho Sui, Victoria~J. Cook, Lena
  Shah, Elizabeth Brodkin, Shirley Rempel, Richard Moore, Yongjun Zhao, Robert
  Holt, Richard Varhol, Inanc Birol, Marcus Lem, Meenu~K. Sharma, Kevin Elwood,
  Steven~J.M. Jones, Fiona~S.L. Brinkman, Robert~C. Brunham, and Patrick Tang.
\newblock Whole-genome sequencing and social-network analysis of a tuberculosis
  outbreak.
\newblock {\em New England Journal of Medicine}, 364(8):730--739, 2011.
\newblock PMID: 21345102.

\bibitem{Morone2015}
Flaviano Morone and Hern{\'{a}}n~A. Makse.
\newblock {Influence maximization in complex networks through optimal
  percolation}.
\newblock {\em Nature}, 524(7563):65--68, 2015.

\bibitem{Valente2012}
Thomas~W. Valente.
\newblock {Network interventions}.
\newblock {\em Science}, 336(6090):49--53, 2012.

\bibitem{10.1239/jap/1389370105}
Frank Ball and David Sirl.
\newblock {Acquaintance vaccination in an epidemic on a random graph with
  specified degree distribution}.
\newblock {\em Journal of Applied Probability}, 50(4):1147 -- 1168, 2013.

\bibitem{Kim2015}
David~A. Kim, Alison~R. Hwong, Derek Stafford, D.~Alex Hughes, A.~James
  O'Malley, James~H. Fowler, and Nicholas~A. Christakis.
\newblock {Social network targeting to maximise population behaviour change: A
  cluster randomised controlled trial}.
\newblock {\em The Lancet}, 386(9989):145--153, 2015.

\bibitem{Chin2021}
Alex Chin, Dean Eckles, and Johan Ugander.
\newblock {Evaluating Stochastic Seeding Strategies in Networks}.
\newblock {\em Management Science}, pages 1--63, 2021.

\bibitem{doi:10.1509/jmkr.48.3.425}
Zsolt Katona, Peter~Pal Zubcsek, and Miklos Sarvary.
\newblock Network effects and personal influences: The diffusion of an online
  social network.
\newblock {\em Journal of Marketing Research}, 48(3):425--443, 2011.

\bibitem{Kempe2003}
David Kempe and Jon Kleinberg.
\newblock {P137-Kempe}.
\newblock {\em Kdd}, pages 137--146, 2003.

\bibitem{Kempe2015}
David Kempe, Jon Kleinberg, and {\'{E}}va Tardos.
\newblock {Maximizing the spread of influence through a social network}.
\newblock {\em Theory of Computing}, 11:105--147, 2015.

\bibitem{Zhu2017}
Jinghua Zhu, Yong Liu, and Xuming Yin.
\newblock {A New Structure-Hole-Based Algorithm for Influence Maximization in
  Large Online Social Networks}.
\newblock {\em IEEE Access}, 5:23405--23412, 2017.

\bibitem{Cohen2003}
Reuven Cohen, Shlomo Havlin, and Daniel Ben-Avraham.
\newblock {Efficient immunization strategies for computer networks and
  populations}.
\newblock {\em Physical Review Letters}, 91(24):1--5, 2003.

\bibitem{Li2016}
Jin Li, Kun Yue, Dehai Zhang, and Weiyi Liu.
\newblock {Robust influence blocking maximization in social networks}.
\newblock {\em Jisuanji Yanjiu yu Fazhan/Computer Research and Development},
  53(3):601--610, 2016.

\bibitem{Goyal2011}
Amit Goyal, Wei Lu, and Laks~V.S. Lakshmanan.
\newblock {CELF++: Optimizing the greedy algorithm for influence maximization
  in social networks}.
\newblock {\em Proceedings of the 20th International Conference Companion on
  World Wide Web, WWW 2011}, pages 47--48, 2011.

\bibitem{NC_complex}
Guilbeault Douglas and Centola Damon.
\newblock Topological measures for identifying and predicting the spread of
  complex contagions.
\newblock {\em Nature Communication}, 12, July 2021.

\bibitem{7899466}
Shodai Mihara, Sho Tsugawa, and Hiroyuki Ohsaki.
\newblock On the effectiveness of random jumps in an influence maximization
  algorithm for unknown graphs.
\newblock In {\em 2017 International Conference on Information Networking
  (ICOIN)}, pages 395--400, 2017.

\bibitem{Mihara2015}
Shodai Mihara, Sho Tsugawa, and Hiroyuki Ohsaki.
\newblock {Influence maximization problem for unknown social networks}.
\newblock {\em Proceedings of the 2015 IEEE/ACM International Conference on
  Advances in Social Networks Analysis and Mining, ASONAM 2015}, pages
  1539--1546, 2015.

\bibitem{Wilder2017}
Bryan Wilder, Nicole Immorlica, Eric Rice, and Milind Tambe.
\newblock {Inuence maximization with an unknown network by exploiting community
  structure}.
\newblock {\em CEUR Workshop Proceedings}, 1893(SocInf):2--7, 2017.

\bibitem{Wilder2018}
Bryan Wilder, Nicole Immorlica, Eric Rice, and Milind Tambe.
\newblock {Maximizing influence in an unknown social network}.
\newblock {\em 32nd AAAI Conference on Artificial Intelligence, AAAI 2018},
  pages 4743--4750, 2018.

\bibitem{Eckles2019}
Dean Eckles, Hossein Esfandiari, Elchanan Mossel, and M.~Amin Rahimian.
\newblock {Seeding with costly network information}.
\newblock {\em ACM EC 2019 - Proceedings of the 2019 ACM Conference on
  Economics and Computation}, 2019:421--422, 2019.

\bibitem{Stein2017}
Sebastian Stein, Soheil Eshghi, Setareh Maghsudi, Leandros Tassiulas,
  Rachel~K.E. Bellamy, and Nicholas~R. Jennings.
\newblock {Heuristic algorithms for influence maximization in partially
  observable social networks}.
\newblock {\em CEUR Workshop Proceedings}, 1893(SocInf):20--32, 2017.

\bibitem{Erkol2017}
Şirag Erkol and G{\"{o}}nen{\c{c}} Y{\"{u}}cel.
\newblock {Influence maximization based on partial network structure
  information: A comparative analysis on seed selection heuristics}.
\newblock {\em International Journal of Modern Physics C}, 28(10), 2017.

\bibitem{Paluck566}
Elizabeth~Levy Paluck, Hana Shepherd, and Peter~M. Aronow.
\newblock Changing climates of conflict: A social network experiment in 56
  schools.
\newblock {\em Proceedings of the National Academy of Sciences},
  113(3):566--571, 2016.

\bibitem{10.2307/2781907}
Scott~L. Feld.
\newblock Why your friends have more friends than you do.
\newblock {\em American Journal of Sociology}, 96(6):1464--1477, 1991.

\bibitem{Chami2017}
Goylette~F. Chami, Sebastian~E. Ahnert, Narcis~B. Kabatereine, and Edridah~M.
  Tukahebwa.
\newblock {Social network fragmentation and community health}.
\newblock {\em Proceedings of the National Academy of Sciences of the United
  States of America}, 114(36):E7425--E7431, 2017.

\bibitem{20.500.12613/4943}
H~Shakya, Stafford Derek, Hughes D, Alex, Keegan Thomas, Negron Rennie, Broome
  Jai, McKnight Mark, Nicoll Liza, Nelson Jennifer, Iriarte Emma, Ordonez
  Maria, Airoldi Edo, H~Fowler James, and Christakis Nicholas, A.
\newblock Exploiting social influence to magnify population-level behaviour
  change in maternal and child health: Study protocol for a randomised
  controlled trial of network targeting algorithms in rural honduras.
\newblock {\em BMJ Open}, 2017.

\bibitem{10.1145/2110363.2110394}
Mohammad~S. Hashemian, Kevin~G. Stanley, Dylan~L. Knowles, Jonathan Calver, and
  Nathaniel~D. Osgood.
\newblock Human network data collection in the wild: The epidemiological
  utility of micro-contact and location data.
\newblock In {\em Proceedings of the 2nd ACM SIGHIT International Health
  Informatics Symposium}, IHI '12, page 255–264, New York, NY, USA, 2012.
  Association for Computing Machinery.

\bibitem{POMARE2019}
Chiara Pomare, Janet~C. Long, Kate Churruca, Louise~A. Ellis, and Jeffrey
  Braithwaite.
\newblock Social network research in health care settings: Design and data
  collection.
\newblock {\em Social Networks}, 2019.

\bibitem{doi:10.1068/b3317t}
Juan~Antonio Carrasco, Bernie Hogan, Barry Wellman, and Eric~J Miller.
\newblock Collecting social network data to study social activity-travel
  behavior: An egocentric approach.
\newblock {\em Environment and Planning B: Planning and Design},
  35(6):961--980, 2008.

\bibitem{5428686}
Stephen~T. Ricken, Richard~P. Schuler, Sukeshini~A. Grandhi, and Quentin Jones.
\newblock Telluswho: Guided social network data collection.
\newblock In {\em 2010 43rd Hawaii International Conference on System
  Sciences}, pages 1--10, 2010.

\bibitem{Akbarpour2017}
Mohammad Akbarpour, Suraj Malladi, and Amin Saberi.
\newblock {Just a Few Seeds More: Value of Network Information for Diffusion}.
\newblock {\em Ssrn}, pages 1--42, 2017.

\bibitem{Lattanzi2015}
Silvio Lattanzi and Yaron Singer.
\newblock {The power of random neighbors in social networks}.
\newblock {\em WSDM 2015 - Proceedings of the 8th ACM International Conference
  on Web Search and Data Mining}, pages 77--86, 2015.

\bibitem{DBLP}
Silvio Lattanzi and Yaron Singer.
\newblock The power of random neighbors in social networks.
\newblock In {\em WSDM}, pages 77--86, 2015.

\bibitem{7990601}
Sancheng Peng, Guojun Wang, Yongmei Zhou, Cong Wan, Cong Wang, Shui Yu, and
  Jianwei Niu.
\newblock An immunization framework for social networks through big data based
  influence modeling.
\newblock {\em IEEE Transactions on Dependable and Secure Computing},
  16(6):984--995, 2019.

\bibitem{ZHANG2018920}
Junbao Zhang, Haojun Huang, Yan Luo, Yinting Fan, and Guan Yang.
\newblock Immunization-based redundancy elimination in mobile opportunistic
  networks-generated big data.
\newblock {\em Future Generation Computer Systems}, 79:920--927, 2018.

\bibitem{centola2007complex}
Damon Centola and Michael Macy.
\newblock Complex contagions and the weakness of long ties.
\newblock {\em American journal of Sociology}, 113(3):702--734, 2007.

\bibitem{centola2010spread}
Damon Centola.
\newblock The spread of behavior in an online social network experiment.
\newblock {\em science}, 329(5996):1194--1197, 2010.

\bibitem{schelling1978micromotives}
Thomas~C. Schelling.
\newblock {\em Micromotives and Macrobehavior}.
\newblock {W. W. Norton \& Company}, October 1978.

\bibitem{valente1995network}
Thomas~W Valente.
\newblock {\em Network models of the diffusion of innovations}.
\newblock Number 303.484 V3. 1995.

\bibitem{watts2002simple}
Duncan~J Watts.
\newblock A simple model of global cascades on random networks.
\newblock {\em Proceedings of the National Academy of Sciences},
  99(9):5766--5771, 2002.

\bibitem{young2020individual}
H~Peyton Young.
\newblock {\em Individual strategy and social structure}.
\newblock Princeton University Press, 2020.

\bibitem{Centola+2018}
Damon Centola.
\newblock {\em How Behavior Spreads}.
\newblock Princeton University Press, 2018.

\bibitem{10.1086/521848}
Damon Centola and Michael Macy.
\newblock Complex contagions and the weakness of long ties.
\newblock {\em American Journal of Sociology}, 113(3):702--734, 2007.

\bibitem{doi:10.1080/10618600.2012.738106}
Stephen~E. Fienberg.
\newblock A brief history of statistical models for network analysis and open
  challenges.
\newblock {\em Journal of Computational and Graphical Statistics},
  21(4):825--839, 2012.

\bibitem{watts_collective_1998}
Duncan~J. Watts and Steven~H. Strogatz.
\newblock Collective dynamics of ‘small-world’ networks.
\newblock {\em Nature}, 393(6684):440--442, 1998.

\bibitem{Albert:2002:rmp}
R\'eka Albert and Albert-L\'aszl\'o Barab\'asi.
\newblock Statistical mechanics of complex networks.
\newblock {\em Reviews of Modern Physics}, 74(1):47--97, January 2002.

\bibitem{albert2000error}
Reka Albert, Hawoong Jeong, and Albert-Laszlo Barabasi.
\newblock Error and attack tolerance of complex networks.
\newblock {\em Nature}, 406(6794):378--382, July 2000.

\bibitem{10.1371/journal.pone.0136497}
Rossana Mastrandrea, Julie Fournet, and Alain Barrat.
\newblock Contact patterns in a high school: A comparison between data
  collected using wearable sensors, contact diaries and friendship surveys.
\newblock {\em PLOS ONE}, 10(9):1--26, 09 2015.

\bibitem{10.1371/journal.pone.0023176}
Juliette Stehlé, Nicolas Voirin, Alain Barrat, Ciro Cattuto, Lorenzo Isella,
  Jean-François Pinton, Marco Quaggiotto, Wouter Van~den Broeck, Corinne
  Régis, Bruno Lina, and Philippe Vanhems.
\newblock High-resolution measurements of face-to-face contact patterns in a
  primary school.
\newblock {\em PLOS ONE}, 6(8):1--13, 08 2011.

\bibitem{ISELLA2011166}
Lorenzo Isella, Juliette Stehlé, Alain Barrat, Ciro Cattuto, Jean-François
  Pinton, and Wouter {Van den Broeck}.
\newblock What's in a crowd? analysis of face-to-face behavioral networks.
\newblock {\em Journal of Theoretical Biology}, 271(1):166--180, 2011.

\bibitem{NIPS2012_7a614fd0}
Jure Leskovec and Julian Mcauley.
\newblock Learning to discover social circles in ego networks.
\newblock In F.~Pereira, C.~J.~C. Burges, L.~Bottou, and K.~Q. Weinberger,
  editors, {\em Advances in Neural Information Processing Systems}, volume~25.
  Curran Associates, Inc., 2012.

\bibitem{leskovec2010signed}
Jure Leskovec, Daniel Huttenlocher, and Jon Kleinberg.
\newblock Signed networks in social media, 2010.

\bibitem{Leskovec2010PredictingPA}
Jure Leskovec, Daniel~P. Huttenlocher, and Jon~M. Kleinberg.
\newblock Predicting positive and negative links in online social networks.
\newblock In {\em WWW '10}, 2010.

\bibitem{kumar2016edge}
Srijan Kumar, Francesca Spezzano, VS~Subrahmanian, and Christos Faloutsos.
\newblock Edge weight prediction in weighted signed networks.
\newblock In {\em Data Mining (ICDM), 2016 IEEE 16th International Conference
  on}, pages 221--230. IEEE, 2016.

\bibitem{kumar2018rev2}
Srijan Kumar, Bryan Hooi, Disha Makhija, Mohit Kumar, Christos Faloutsos, and
  VS~Subrahmanian.
\newblock Rev2: Fraudulent user prediction in rating platforms.
\newblock In {\em Proceedings of the Eleventh ACM International Conference on
  Web Search and Data Mining}, pages 333--341. ACM, 2018.

\bibitem{feather}
Benedek Rozemberczki and Rik Sarkar.
\newblock {Characteristic Functions on Graphs: Birds of a Feather, from
  Statistical Descriptors to Parametric Models}.
\newblock In {\em Proceedings of the 29th ACM International Conference on
  Information and Knowledge Management (CIKM '20)}, page 1325–1334. ACM,
  2020.

\bibitem{10.1145/1217299.1217301}
Jure Leskovec, Jon Kleinberg, and Christos Faloutsos.
\newblock Graph evolution: Densification and shrinking diameters.
\newblock {\em ACM Trans. Knowl. Discov. Data}, 1(1):2–es, March 2007.

\bibitem{article}
Matei Ripeanu, Ian Foster, and Adriana Iamnitchi.
\newblock Mapping the gnutella network: Properties of large-scale peer-to-peer
  systems and implications for system design.
\newblock {\em IEEE Internet Computing Journal}, 6, 10 2002.

\bibitem{10.1145/3097983.3098069}
Hao Yin, Austin~R. Benson, Jure Leskovec, and David~F. Gleich.
\newblock Local higher-order graph clustering.
\newblock In {\em Proceedings of the 23rd ACM SIGKDD International Conference
  on Knowledge Discovery and Data Mining}, KDD '17, page 555–564, New York,
  NY, USA, 2017. Association for Computing Machinery.

\bibitem{Morries2000}
S.~Morris.
\newblock Contagion.
\newblock {\em The Review of Economic Studies}, 67:57--78, 2000.

\bibitem{Berger2001}
E.~Berger.
\newblock Dynamic monopolies of constant size.
\newblock {\em Journal of Combinatorial Theory}, 83:191--200, 2001.

\bibitem{Crucitti2004ErrorAA}
Paolo Crucitti, Vito Latora, Massimo Marchiori, and Andrea Rapisarda.
\newblock Error and attack tolerance of complex networks.
\newblock {\em Physica A-statistical Mechanics and Its Applications},
  340:388--394, 2004.

\bibitem{Tsiotas6701}
Dimitrios Tsiotas.
\newblock Detecting different topologies immanent in scale-free networks with
  the same degree distribution.
\newblock {\em Proceedings of the National Academy of Sciences},
  116(14):6701--6706, 2019.

\bibitem{Kumar2019}
Vineet Kumar and K.~Sudhir.
\newblock {Can Friends Seed More Buzz and Adoption ?}
\newblock {\em Cowles Foundation Discussion Paper}, 2178R(2178), 2021.

\bibitem{Kitsak2010}
Maksim Kitsak, Lazaros~K. Gallos, Shlomo Havlin, Fredrik Liljeros, Lev Muchnik,
  H.~Eugene Stanley, and Hern{\'{a}}n~A. Makse.
\newblock {Identification of influential spreaders in complex networks}.
\newblock {\em Nature Physics}, 6(11):888--893, 2010.

\end{thebibliography}

\newpage
\section{Supplementary Plots}
\renewcommand{\thefigure}{S\arabic{figure}}

\begin{figure}[!h]
    \centering
    \includegraphics[width=1\linewidth]{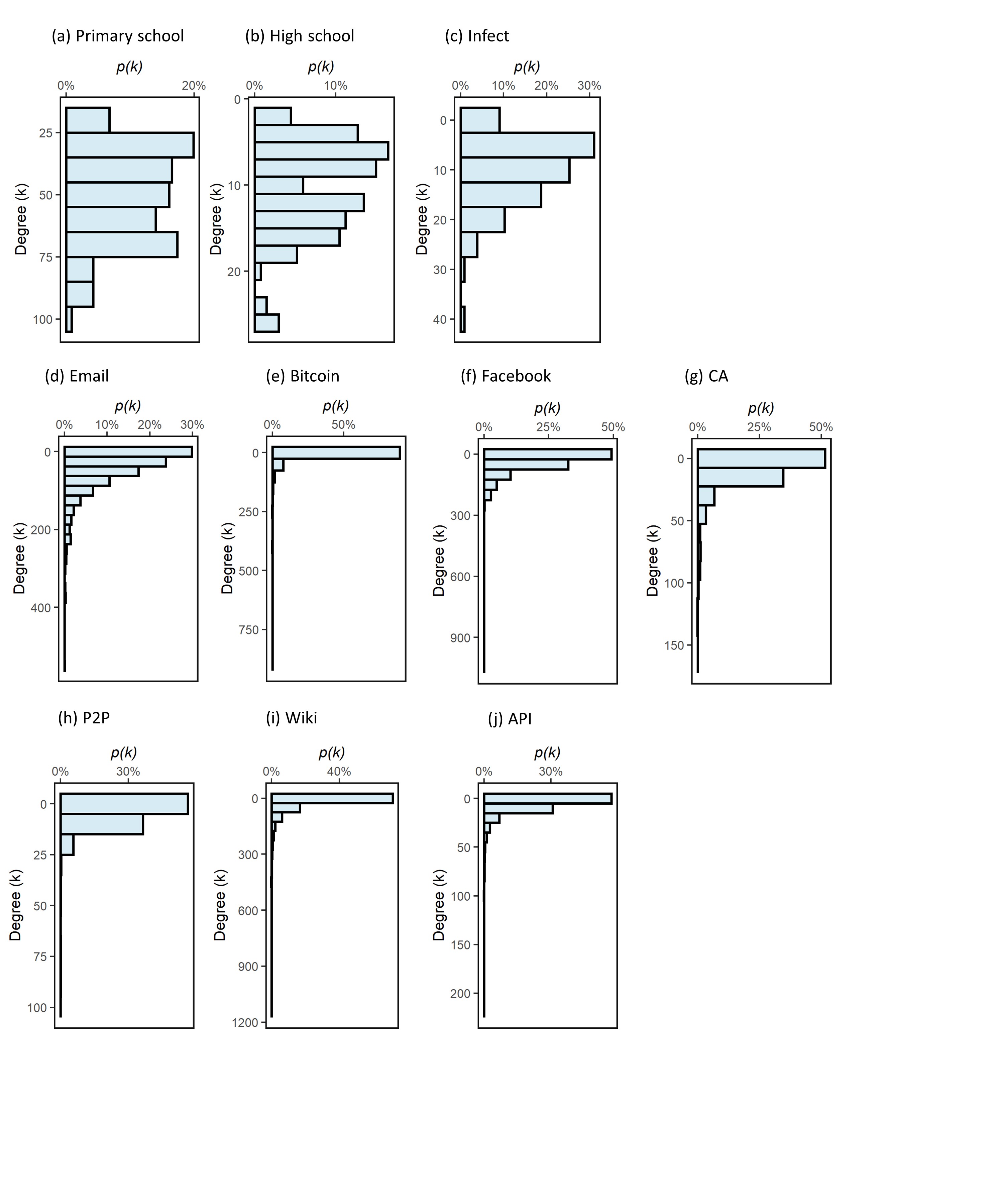}
    \caption{Degree distributions of all the social networks used in this study. (a-c) are offline social networks for a primary school, a high school and an exhibition (Infect), respectively. (d-j) are online social networks for email communication (Email), Bitcoin trading, Facebook, academic author collaboration (CA), Gnutella peer to peer (p2p), Wiki users, and LastFMsuers from Asia (API), respectively. Degree variances in online social networks are much higher.}
    \label{figS1}
\end{figure}

\begin{figure}[!h]
    \centering
    \includegraphics[width=1\linewidth]{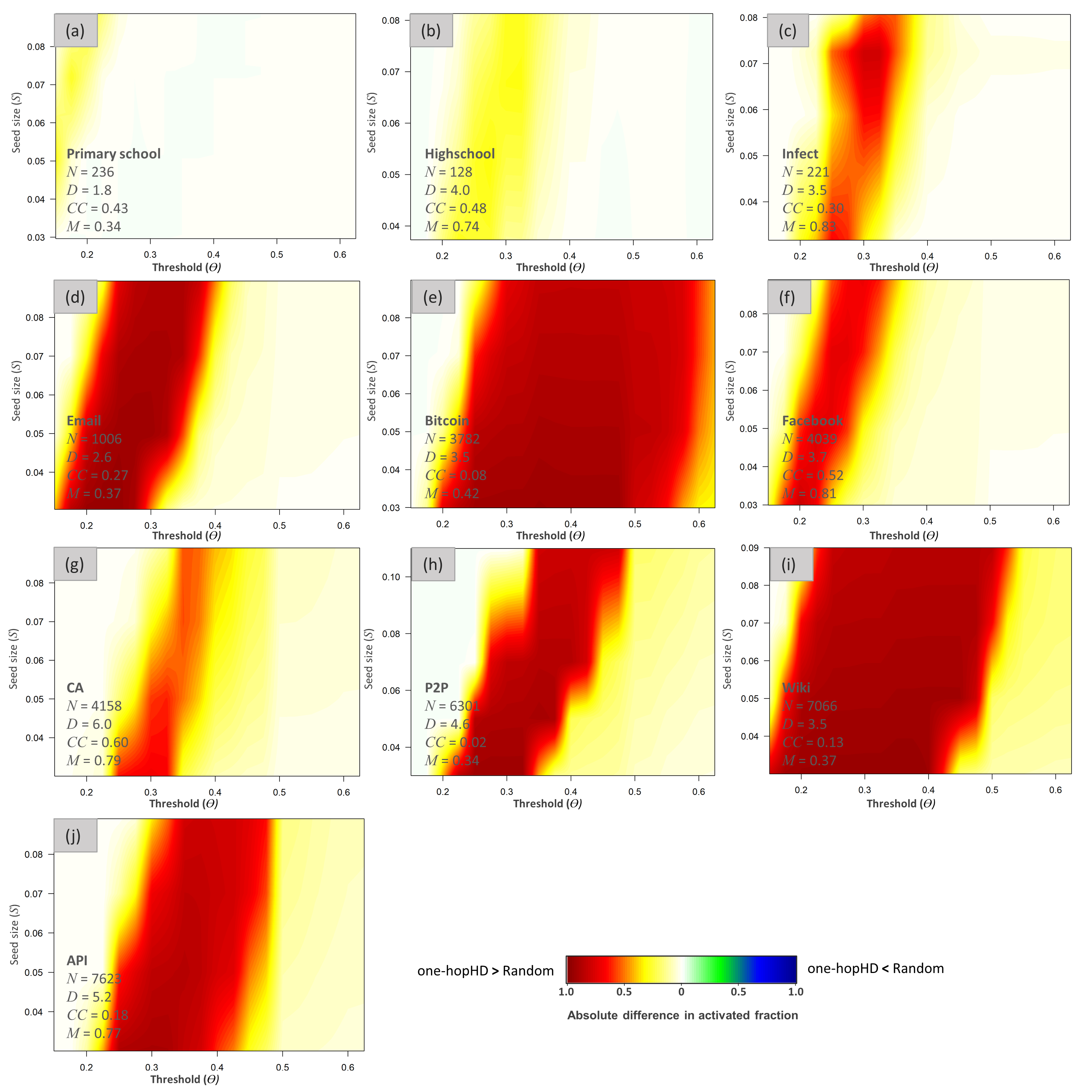}
    \caption{Comparison of the final activated fractions between one-hopHD and random seeding in 10 real networks (a-j), with same number of seeds. The x-axis is the value of uniform threshold used in the contagion model, and the y-axis is the seed size. The color gradient in (a-j) represents the absolute difference in terms of the final activated fractions by one-hopHD and random strategies (average of 10,000 runs), under specific combination of threshold and seed set size. The coordinates of every color pixel corresponds to threshold (value of x-axis) and seed set size (value of y-axis), respectively. With the same number of seeds, one-hopHD leads to higher final activated fraction than random strategy in all the networks tested here.}
    \label{figS2}
\end{figure}

\end{document}